\documentclass[showpacs,aps,prd,nofootinbib,floatfix,amsmath,amssymb]{revtex4}
\usepackage{graphicx}
\usepackage[utf8x]{inputenc}

\begin{document}

\makeatletter
\newbox\slashbox \setbox\slashbox=\hbox{$/$}
\newbox\Slashbox \setbox\Slashbox=\hbox{\large$/$}
\def\pFMslash#1{\setbox\@tempboxa=\hbox{$#1$}
  \@tempdima=0.5\wd\slashbox \advance\@tempdima 0.5\wd\@tempboxa
  \copy\slashbox \kern-\@tempdima \box\@tempboxa}
\def\pFMSlash#1{\setbox\@tempboxa=\hbox{$#1$}
  \@tempdima=0.5\wd\Slashbox \advance\@tempdima 0.5\wd\@tempboxa
  \copy\Slashbox \kern-\@tempdima \box\@tempboxa}
\def\FMslash{\protect\pFMslash}
\def\FMSlash{\protect\pFMSlash}
\def\miss#1{\ifmmode{/\mkern-11mu #1}\else{${/\mkern-11mu #1}$}\fi}
\makeatother

\title{Trilinear gauge boson couplings in the standard model with one universal extra
dimension}
\author{M. A. L\'opez--Osorio$^{(a)}$, E. Mart\'inez--Pascual$^{(a)}$, J. Monta\~no$^{(b)}$, H. Novales--S\'anchez$^{( c )}$, J. J. Toscano$^{(a)}$, and E. S. Tututi$^{( c )}$}
\address{$^{(a)}$ Facultad de Ciencias F\'isico Matem\'aticas, Benem\'erita Universidad Aut\'onoma de Puebla, Apartado Postal 1152, Puebla, Puebla, M\'exico.\\ $^{(b)}$ Departamento de F\'isica, CINVESTAV, Apartado Postal 14-740, 07000, M\'exico, D. F., M\'exico.\\ $^{( c )}$ Facultad de Ciencias F\'isico Matem\'aticas, Universidad Michoacana de San Nicol\'as de Hidalgo, Avenida Francisco J. M\'ujica S/N, 58060, Morelia Michoac\'an, M\'exico.}
\begin{abstract}
One--loop effects of Standard Model (SM) extensions comprising universal extra dimensions are essential as a consequence of Kaluza--Klein (KK) parity conservation, for they represent the very first presumable virtual effects on low--energy observables. In this paper, we calculate the one--loop $CP$--even contributions to the SM $WW\gamma$ and $WWZ$ gauge couplings produced by the KK excited modes that stand for the dynamical variables of the effective theory emerged from a generalization of the SM to five dimensions, in which the extra dimension is assumed to be universal, after compactification. The employment of a covariant gauge--fixing procedure that removes gauge invariance associated to gauge KK excited modes, while keeping electroweak gauge symmetry manifest, is a main feature of this calculation, which is performed in the Feynman 't Hooft gauge and yields finite results that consistently decouple for a large compactification scale. After numerical evaluation, our results show to be comparable with the one--loop SM contributions and well within the reach of a next linear collider.
\end{abstract}


\maketitle

\section{Introduction}
\label{intro}
The interesting idea that space--time comprises a larger number of spatial dimensions than those that we have been able to detect, even with the aid of the most powerful colliders running right now, inspires attractive extensions of the Standard Model (SM) that pursue the fundamental theory describing nature at high energies. The first models involving extra spatial dimensions were conceived long ago~\cite{TKOK}, but this sort of descriptions gained remarkable attention just a few years ago, when it was pointed out~\cite{AADD} that extra dimensions could be found at the TeV scale. In the present paper, we consider a generalization of the SM to five dimensions under the assumption that the extra dimension is {\it universal}~\cite{ACD}, which means that all fields in the model propagate in the bulk. Models involving universal extra dimensions (UED) possess physical interest, as it is exhibited in various works covering areas of high energy physics such as dark matter~\cite{UEDdm}, neutrino physics~\cite{UEDn}, Higgs physics~\cite{UEDh}, and flavor physics~\cite{UEDf}. Up--to--date we have not found any indication that extra dimensions actually exist, yet a Kaluza--Klein type of effective theory is not in contradiction with current experimental data. Experimental consistency of this sort of effective theories is ensured by the assumption that spatial extra dimensions are compactified, since compact extra dimensions would have remained so far out of the reach of our most sensitive experiments if they are small enough.

A striking consequence of compact extra dimensions is the inception of an infinite set of fields, defined on four--dimensional Minkowski spacetime, in addition to the SM ones. There are different compactification schemes, and the consideration of more than one extra dimension comes along with more options to do it. In the case of one extra dimension, the orbifold compactification on $S^1/Z_2$, which is the simplest approach that allows one to reproduce the SM as the low--energy physical description, yields periodicity and parity properties of the dynamical variables with respect to the extra--dimensional coordinate. Then the geometry of the orbifold opens the possibility of expanding the ``fundamental'' fields in Fourier series, known as Kaluza--Klein (KK) towers. Each term of these series incorporates a field, a KK mode, propagating in the usual four--dimensional space--time. KK modes can be classified into zero modes and excited modes, depending whether they are the zero mode in the Fourier expansion or not. Zero modes are identified with light fields, that is, the SM's dynamical variables, and KK excited modes are associated with new states whose presence has not been noticed by experiments so far. In every Fourier expansion of fundamental fields the whole dependence on the extra--dimension is collected within trigonometric functions; integrating out the extra--dimensional coordinate in the action provides an effective four--dimensional Lagrangian where the KK modes enter as effective dynamical variables. Gauge parameters defining the extra--dimensional gauge transformations are, according to the field--antifield formalism~\cite{GPS}, dynamical variables, as they are made to coincide with ghost fields, that is such parameters propagate in the bulk and can be expanded into KK towers~\cite{NT1}.

After expanding gauge covariant objects in the higher dimensional theory, the integration of the compact extra dimension provides an effective Lagrangian that is invariant under the standard gauge transformations (SGT) and the non-standard gauge transformations (NSGT), both with gauge parameters defined on the four--dimensional Minkowski spacetime~\cite{NT1}. In a more recent work~\cite{LMNT}, it was shown (using pure Yang--Mills theory) that the KK expansions of extra--dimensional gauge fields define a point transformation that connects the fundamental theory and the effective one, in such a way that objects with well defined transformation laws under the gauge group of the former are mapped into objects with well defined transformation laws under the SGT present in latter. At a phase space level, this transformation can be lift to a  canonical transformation. Fourier expansions of gauge parameters that propagate in the bulk show that this mapping sends the extra--dimensional gauge group into two disjoint subsets. One of them is the set of SGT, which forms a group exclusively defined by the zero
modes of the gauge parameters, and the other is the set of NSGT, which in contrast does not
form a group. It is in this sense that  the full extra--dimensional gauge symmetry is kept nontrivially hidden within the KK theory.

Employing the concepts of SGT and NSGT, the effective KK theory obtained from the five--dimensional pure $ SU_{5}(N) $ Yang--Mills theory\footnote{Also referred to as  pure $ SU(N,\mathcal{M}^{5}) $ Yang-Mills~\cite{LMNT}}  was quantized within the field--antifield framework~\cite{NT1}. In the same fashion, the quantization of the whole five--dimensional Standard Model (5DSM), in the UED context, was performed~\cite{CGNT}. Each set of gauge transformations, the  SGT and NSGT, are characterized by gauge parameters independent of each other, hence the SGT and NSGT independently leave invariant the SM with one UED. One may wish to fix a gauge involving only the NSGT in a SGT invariant way\footnote{SGT can also be fixed, eliminating all gauge invariance from the theory. However, the goals of the present paper do not require it.}; in Ref.~\cite{CGNT} such covariant gauge--fixing procedure was provided. The resulting tree--level structure of the effective theory was examined, including a comprehensive list of expressions of tree--level interactions and the appropriate definitions of all mass eigenstates.

As there exist high--energy phenomena not described by the SM, the existence of extensions is well motivated. A way in which new physics may manifest is through the $WWV$ interactions, with
V representing a SM neutral gauge boson. These trilinear gauge couplings (TGC's), which have been studied in different contexts such as supersymmetry~\cite{TGCsusy}, extra dimensions~\cite{FMNRT}, and extensions to the SM gauge group~\cite{MTTR}, offer the possibility of finding evidence of this new physics at high--energy through virtual effects on SM observables. At this point, the following crucial feature of UED plays a role: The very first contributions of this sort of SM extensions to low--energy Green's functions is at  one--loop level, as no tree--level effects on them exist~\cite{ACD}. This is an implication of the so--called KK--parity conservation, which is an exclusive attribute of models involving UED that makes the bounds on the size of the involved extra dimensions relatively weak. The importance of one--loop corrections to SM observables from models with UED relies not only on this issue, as the renormalizability~\cite{NT1,NT2} of contributions at this order provides the possibility of obtaining unambiguous results, even in spite of the well--known nonrenormalizable comportment supplied by the the presence of dimensionful coupling constants in extra--dimensional models.

In the present paper, we use the gauge fixing procedure and results reported in Ref.~\cite{CGNT} to derive the $CP$-even contributions to the TGC's $WWV$.  The fact that the only new parameter introduced by UED models is the size of the extra dimension, enhances the predictive power of these kind of models and simultaneously becomes an incentive to perform this calculation. We take the $W$ bosons on shell, but leave the neutral gauge boson off shell. We derive, in the Feynman--'t Hooft gauge, the anomalous contributions to the form factors parametrizing these interactions and find finite results. Then, as an interesting case, we consider the heavy--compactification scenario, which we define by the condition $Q^2<<1/R^2$, where $Q$ is the momentum of the external neutral gauge boson and $R$ is the radius of the orbifold--compactified extra dimension. The new--physics contributions, which we formerly expressed in terms of the Passarino--Veltman scalar functions~\cite{PV}, are exhibited as elementary functions of masses and the compactification scale. The numerical evaluation of the derived expressions then gives an estimation of extra--dimensional effects on the TGC's of interest to the present paper for a linear collider with a center--of--mass energy $E_{\rm CM}=\sqrt{Q^2}=500$~GeV and different compactification scales, ranging from $1$~TeV to $3$~TeV.

The organization of this paper is as follows. In Section \ref{EDSM} we provide some necessary information about the SM with one UED and the KK theory that it generates after compactification has taken place. Then, in Section \ref{EDWWV}, we describe our calculation of the $CP$--even anomalous contributions to the $WWV$ TGC's and consider the heavy--compactification scenario. This section also includes a numerical estimation of the extra--dimensional effects on this interaction and a discussion of results. Our conclusions are presented in Section \ref{conc}. Finally, we include an Appendix were the Lagrangian terms contributing to the $ WWV $ vertex at the one--loop  level are provided.

\section{The Standard Model with one universal extra dimension}
\label{EDSM}
In this section we define our notation and provide a general description of the context within which we shall perform the phenomenological calculations. We define the five--dimensional model and introduce the KK expansions used in order to preserve gauge invariance in the transit from five to four dimensions. We then briefly discuss the scheme to covariantly fix the gauge and supply the corresponding gauge--fixing functions. Finally, we give the appropriate transformations that set all mass eigenstates. As all these ideas have been addressed extensively in Refs.~\cite{NT1,CGNT,LMNT}, our discussion is intended to be succinct.

\subsection{The five--dimensional model}
Consider a five--dimensional Minkowski spacetime, with mostly negative metric, on which the following Lagrangian is defined
\begin{equation}
{\cal L}_{\rm 5DSM}(x,y)={\cal L}_{\rm 5DG}(x,y)+{\cal L}_{\rm 5DH}(x,y)+{\cal L}_{\rm 5DC}(x,y)+{\cal L}_{\rm 5DY}(x,y);
\end{equation}
where all dynamical variables are defined on the five dimensional spacetime. Ordinary four--dimensional coordinates have been denoted by $x$ whereas the fifth dimension is labeled by $y$. This five--dimensional Lagrangian, whose gauge group is $SU_5(3)_{\rm C}\times SU_5(2)_W\times U_5(1)_Y$, is composed by the Gauge (${\cal L}_{\rm 5DG}$), Higgs ( ${\cal L}_{\rm 5DH}$) , Currents (${\cal L}_{\rm 5DC}$) and Yukawa (${\cal L}_{\rm 5DY}$) sectors.

 The Gauge sector is defined as
\begin{equation}
{\cal L}_{\rm 5DG}(x,y)=-\frac{1}{4}{\cal G}^a_{MN}(x,y){\cal G}^{aMN}(x,y)-\frac{1}{4}{\cal W}^i_{MN}(x,y){\cal W}^{iMN}(x,y)-\frac{1}{4}{\cal B}_{MN}(x,y){\cal B}^{MN}(x,y).
\end{equation}
In this expression and throughout the rest of the paper, Lorentz indices are denoted by $M,N,\ldots=0,1,2,3,5$ and gauge indices corresponding to $SU_5(3)_{\rm C}$ and $SU_5(2)_W$ are denoted by $a,b,c,\ldots$ and $i,j,k,\ldots$, respectively. In addition, greek indices $\mu,\nu,\ldots=0,1,2,3$ will label four--dimensional Lorentz indices. Five--dimensional field strengths, ${\cal G}^a_{MN}$, ${\cal W}^i_{MN}$ and ${\cal B}_{MN}$, are defined by
\begin{eqnarray}
{\cal G}^a_{MN}(x,y)&=&\partial_M{\cal G}^a_N(x,y)-\partial_N{\cal G}^a_M(x,y)+g^{\rm s}_5f^{abc}{\cal G}^b_M(x,y){\cal G}^c_N(x,y),
\\ \nonumber \\
{\cal W}^i_{MN}(x,y)&=&\partial_M{\cal W}^i_N(x,y)-\partial_N{\cal W}^i_M(x,y)+g_5\epsilon^{ijk}{\cal W}^j_M(x,y){\cal W}^k_N(x,y),
\\ \nonumber \\
{\cal B}_{MN}(x,y)&=&\partial_M{\cal B}_N(x,y)-\partial_N{\cal B}_M(x,y),
\end{eqnarray}
where ${\cal G}^a_M$,  ${\cal W}^i_M$ and ${\cal B}_M$ represent gauge fields for $SU_5(3)_{\rm C}$,  $SU_5(2)_W$ and $U_5(1)_Y$ gauge groups, respectively. In addition, $g^{\rm s}_5$ and $g_5$ are the $SU_5(3)_{\rm C}$ and $SU_5(2)_W$ constant couplings, respectively, both with $({\rm mass})^{-1/2}$ units. Structure constants $f^{abc}$ define the Lie algebra of $SU_5(3)_{\rm C}$, whereas $\epsilon^{ijk}$ (Levi--Civita symbol) defines the Lie algebra of $SU_5(2)_W$. Notice that due to five--dimensional Lorentz symmetry, all vectorial gauge fields are built from five scalar fields.

The gauge symmetry group $SU_5(3)_{\rm C}\times SU_5(2)_W\times U_5(1)_Y$ determines the following infinitesimal transformation laws:
\begin{eqnarray}
\delta{\cal G}^a_M(x,y)&=&\tilde{{\cal D}}^{ab}_M\beta^b(x,y),
\\ \nonumber \\
\delta{\cal W}^i_M(x,y)&=&{\cal D}^{ij}_M\alpha^j(x,y),
\\ \nonumber \\
\delta{\cal B}_M(x,y)&=&\partial_M\alpha(x,y),
\end{eqnarray}
where $\tilde{\cal D}^{ab}_M=\delta^{ab}\partial_M-g^{\rm s}_5f^{abc}{\cal G}^c_M$ is the $SU_5(3)_{\rm C}$ covariant derivative and ${\cal D}^{ij}_M=\delta^{ij}\partial_M-g_5\epsilon^{ijk}{\cal W}^k_M$ plays the analogous role for the $SU_5(2)_W$ group. The most general infinitesimal gauge transformations are defined by requiring gauge parameters to propagate in the bulk. In the field--antifield formalism~\cite{GPS}, these gauge parameters are recognized as ghost fields, which, in the UED framework, also propagate in the extra dimension.

The Higgs sector is given by
\begin{equation}
{\cal L}_{\rm 5DH}(x,y)=(d_M\Phi)^\dag(x,y)(d^M\Phi)(x,y)-\mu^2\Phi^\dag(x,y)\Phi(x,y)-\lambda_5\left[ \Phi^\dag(x,y)\Phi(x,y) \right]^2,
\end{equation}
where $\mu$ has units of mass, whereas the units of $\lambda_5$ are $({\rm mass})^{-1}$. The covariant derivative acting on the Higgs doublet $\Phi(x,y)$, with hypercharge $Y=1$, is given by
\begin{equation}
d_M=\partial_M-ig_5\frac{\sigma^i}{2}{\cal W}^i_M-ig'_5\frac{Y}{2}{\cal B}_M,
\end{equation}
with $g'_5$ representing the $U_5(1)_Y$ coupling constant and $\sigma^i$ standing for the Pauli matrices.

The Currents sector reads
\begin{eqnarray}
{\cal L}_{\rm 5DC}(x,y)&=&\sum_{L_e,L_\mu,L_\tau}i\bar{L}(x,y)\Gamma^M\,(D_ML)(x,y)
+\sum_{Q^u_d,Q^c_s,Q^t_b}i\bar{Q}(x,y)\Gamma^M\,(D_MQ)(x,y) \nonumber
\\  &&
+\sum_{e,\mu,\tau}i\bar{l}(x,y)\Gamma^M\,(D_Ml)(x,y)
+\sum_{u,c,t}i\bar{u}(x,y)\Gamma^M\,(D_Mu)(x,y)+\sum_{d,s,b}i\bar{d}(x,y)\Gamma^M\,(D_Md)(x,y),
\end{eqnarray}
where $\Gamma^M=\gamma^\mu,i\gamma_5$, so that the Clifford algebra $\Gamma^M\Gamma^N+\Gamma^N\Gamma^M=2g^{MN}$, with $g_{MN}={\rm diag}(1,-1,-1,-1,-1)$, is satisfied. In this equation, $l$ is an $SU_5(2)_W$ singlet which collectively denotes the five--dimensional leptonic fields $e$, $\mu$, and $\tau$. There are also up and down five--dimensional quarks $SU_5(2)_W$ singlets, which are represented by $u$ and $d$. On the other hand, $L$ and $Q$ are fermionic $SU_5(2)_W$ doublets corresponding to leptons and quarks, respectively. These fermionic doublets are arranged as  follows:
\begin{equation}
L_l=\left(
\begin{array}{c}
\nu_l \\  l
\end{array}\right),\hspace{1cm}Q^u_d=\left(
\begin{array}{c}
u \\ d
\end{array}\right).
\end{equation}
Finally, the covariant derivative, $D_M$, is defined as
\begin{equation}
D_M=\partial_M-ig_5^{\rm s}\frac{\lambda^a}{2}{\cal G}^a_M-ig_5\frac{\sigma^i}{2}{\cal W}^i_M-ig'_5\frac{Y}{2}{\cal B}_M,
\end{equation}
where $\lambda^a$ are the Gell--Mann matrices.

The term for the Yukawa sector is
\begin{eqnarray}
{\cal L}_{\rm 5DY}(x,y)&=&\sum_{\rm families}\left[\lambda_5^l\bar{L}(x,y)l(x,y)\Phi(x,y)+{\rm h.c.}\right]+\sum_{\rm families}\left[ \lambda_5^u\bar{Q}^u_d(x,y)u(x,y)\tilde{\Phi}(x,y)+{\rm h.c.}\right] \nonumber
\\ &&
+\sum_{\rm families}\left[ \lambda_5^d\bar{Q}^u_d(x,y)d(x,y)\Phi(x,y)+{\rm h.c.}\right],
\end{eqnarray}
in which the couplings $\lambda_5^{l,u,d}$ are dimensionful with units of $({\rm mass})^{-1/2}$. Besides, we have defined $\tilde{\Phi}(x,y)=i\sigma^2\Phi^*(x,y)$.

We now proceed to compactify and integrate out the fifth dimension in order to obtain a four--dimensional KK theory. The effective theory so obtained comprises the four--dimensional SM and a rich variety of new interactions. In the compactification process the full--dimensional gauge invariance will be hidden into the SGT and NSGT. As this work is aimed to derive corrections to electroweak interactions, henceforth those interactions involving $SU_5(3)_{\rm C}$ gauge fields will be disregarded.

\subsection{Compactification and the four--dimensional dynamical variables}

We will study the 5DSM where the extra dimension is compactified on the $S^1/Z_2$--orbifold. We denote the radius of $ S^{1} $ as $R$. This compactification means that one has periodicity on the extra dimension that will allow us to expand in Fourier series fields, covariant objects and gauge parameters of the theory; all of which will collectively be denoted by  $\varphi(x,y)$. Therefore one has $\varphi(x,y)=\varphi(x,y+2\pi R)$ and the following expansion
\begin{equation}
\label{gKKt}
\varphi(x,y)=\frac{1}{\sqrt{2\pi R}}\,\varphi^{(0)}_{\rm even}(x)+\sum_{n=1}^\infty\left[ \frac{1}{\sqrt{\pi R}}\,\varphi^{(n)}_{\rm even}(x)\,{\rm cos}\left( \frac{ny}{R} \right)+\frac{1}{\sqrt{\pi R}}\,\varphi^{(n)}_{\rm odd}(x)\,{\rm sin}\left( \frac{ny}{R} \right) \right],
\end{equation}
where superscripts within parentheses label Fourier modes. This expansion is known as KK tower. Four--dimensional object $\varphi^{(0)}_{\rm even}(x)$, is known as KK zero mode. When $ \varphi $ represents a gauge field, this zero mode is regarded as a low--energy dynamical variable. Each KK tower also involves an infinite set of four--dimensional functions denoted by $\varphi^{(n)}_{\rm even}(x)$ and $\varphi^{(n)}_{\rm odd}(x)$, that are referred to as KK excited modes. For case of fundamental fields, these correspond to heavy fields. Our compactification choice on $S^1/Z_2$ involves a $Z_2$ symmetry, therefore one may conveniently assume defined parity properties on the dynamical variables with respect to reflections $ y\to -y $. Note from Eq.(\ref{gKKt}), that even functions with respect to $ y $ will only be mapped to four--dimensional functions $\varphi^{(0)}_{\rm even}$ and $\varphi^{(n)}_{\rm even}$; by contrast to the case of odd functions with respect to $ y $ which only yield four--dimensional objects $\varphi^{(n)}_{\rm odd}$. We will assume that the fundamental gauge fields and the Higgs doublet have the following parity properties and KK expansions:
\vspace*{.2cm}

{\it Gauge fields}
\begin{eqnarray}
{\cal W}^i_\mu(x,y)=&{\cal W}^i_\mu(x,-y),\hspace{1cm}&{\cal W}^i_\mu(x,y)=\frac{1}{\sqrt{2\pi R}}W^{(0)i}_\mu(x)+\sum_{n=1}^\infty\frac{1}{\sqrt{\pi R}}W^{(n)i}_\mu(x)\,{\rm cos}\left( \frac{ny}{R} \right)
\\ \,\nonumber \\
{\cal W}^i_5(x,y)=&-{\cal W}^i_5(x,-y),\hspace{1cm}&{\cal W}^i_5(x,y)=\sum_{n=1}^\infty\frac{1}{\sqrt{\pi R}}W^{(n)i}_5(x)\,{\rm sin}\left( \frac{ny}{R} \right)
\\ \,\nonumber \\
{\cal B}_\mu(x,y)=&{\cal B}_\mu(x,-y),\hspace{1cm}&{\cal B}_\mu(x,y)=\frac{1}{\sqrt{2\pi R}}B^{(0)}_\mu(x)+\sum_{n=1}^\infty\frac{1}{\sqrt{\pi R}}B^{(n)}_\mu(x)\,{\rm cos}\left( \frac{ny}{R} \right)
\\ \,\nonumber \\
{\cal B}_5(x,y)=&-{\cal B}_5(x,-y),\hspace{1cm}&{\cal B}_5(x,y)=\sum_{n=1}^\infty\frac{1}{\sqrt{\pi R}}B^{(n)}_5(x)\,{\rm sin}\left( \frac{ny}{R} \right)
\end{eqnarray}
{\it Higgs doublet}
\begin{eqnarray}
\Phi(x,y)=&\Phi(x,-y),\hspace{1cm}&\Phi(x,y)=\frac{1}{\sqrt{2\pi R}}\Phi^{(0)}(x)+\sum_{n=1}^\infty\frac{1}{\sqrt{\pi R}}\Phi^{(n)}(x)\,{\rm cos}\left( \frac{ny}{R} \right)
\end{eqnarray}
Notice that choosing either even or odd parity is arbitrary and it is a matter of convenience. For instance, in Gauge--Higgs unification scenarios~\cite{GHu} it is customary to pick even parity for gauge fields with $ M=5 $ so that the resulting KK towers contain KK zero modes. However, we eliminate such four--dimensional degrees of freedom by choosing five--dimensional gauge fields to be even for $ M=\mu $ but to be odd for $ M=5 $.

In five--dimensional frameworks, in which chirality is absent, the $S^1/Z_2$--orbifold compactification allows one to obtain four--dimensional fermionic chiral states and to define $SU_4(2)_{\rm L}$ gauge fields distinguishing chirality, as it occurs in the SM. Five--dimensional Dirac fields, $\psi(x,y)$, transform under parity with respect to the extra dimension as $\psi(x,y)\to\gamma^5\psi(x,-y)$, so we assume the following transformation properties for $SU_5(2)_W$--singlets, $f(x,y)$, and $SU_5(2)_W$--doublets, $F(x,y)$:
\vspace*{.2cm}

{\it Fermionic singlets}
\begin{eqnarray}
\nonumber
f(x,y)\to\gamma^5f(x,-y)&=&f(x,y),
\\ \nonumber \\
f(x,y)&=&\frac{1}{\sqrt{2\pi R}}f^{(0)}_{\rm R}(x)+\sum_{n=1}^\infty\frac{1}{\sqrt{\pi R}}\left[ \hat{f}^{(n)}_{\rm R}(x)\,{\rm cos}\left( \frac{ny}{R} \right)+\hat{f}^{(n)}_{\rm L}(x)\,{\rm sin}\left( \frac{ny}{R} \right) \right]
\end{eqnarray}
{\it Fermionic doublets}
\begin{eqnarray}
\nonumber
F(x,y)\to\gamma^5F(x,-y)&=&-F(x,y),
\\ \nonumber \\
F(x,y)&=&\frac{1}{\sqrt{2\pi R}}F^{(0)}_{\rm L}(x)+\sum_{n=1}^\infty\frac{1}{\sqrt{\pi R}}\left[ F^{(n)}_{\rm L}(x)\,{\rm cos}\left( \frac{ny}{R} \right)+F^{(n)}_{\rm R}(x)\,{\rm sin}\left( \frac{ny}{R} \right) \right]
\end{eqnarray}
where the KK towers are also shown. We have defined $f^{(0)}_{\rm R,L}=(1/2)(1\pm\gamma^5)f^{(0)}$, and analogous definitions for the KK excited modes $f^{(n)}_{\rm R}$, $f^{(n)}_{\rm L}$, $\hat{f}^{(n)}_{\rm R}$ and $\hat{f}^{(n)}_{\rm L}$ hold. These transformation properties suitably generate, at the four--dimensional level, only right--handed singlets and left--handed doublets.

\subsection{Preserving gauge invariance in the Kaluza--Klein theory}
As one can appreciate in Eq.(\ref{gKKt}), all the dependence on $ y $ of covariant objects and gauge parameters is situated in trigonometric functions, hence  one can straightforwardly integrate out the extra--dimensional coordinate in the fundamental action to obtain the effective action
\begin{equation}
S_{\rm eff}=\int d^4x\,{\cal L}_{\rm KK}(x)
\end{equation}
The gauge structure of the fundamental theory determines the gauge structure of $ {\cal L}_{\rm KK} $. As it can be seen from Eq. (\ref{gKKt}), at the four--dimensional level there is an infinite number of gauge parameters, each one of them defining a gauge transformation.

 As it was pointed out in Refs.~\cite{NT1,LMNT}, and implemented to the whole 5DSM in Ref.~\cite{CGNT}, one obtains a KK theory invariant under an infinite number of gauge transformations by expanding five--dimensional covariant objects. Such expansions engender four--dimensional structures that can be fairly considered as gauge--covariant objects. As mentioned in the Introduction, the gauge transformations of the effective KK theory can be divided into two types. One of them, the SGT, consistently coincides with the usual SM variations. Explicitly, the KK modes supplied from the five--dimensional gauge and Higgs sectors transform under the SGT as
\begin{align}
\delta_{\rm s} W^{(0)i}_\mu & =\ {\cal D}^{(0)ij}_\mu\alpha^{(0)j},\quad
\delta_{\rm s} W^{(n)i}_\mu = g\epsilon^{ijk}W^{(n)j}_\mu\alpha^{(0)k},\quad
\delta_{\rm s} W^{(n)i}_5=g\epsilon^{ijk}W^{(n)j}_5\alpha^{(0)k},\\
\delta_{\rm s} B^{(0)}_\mu & =\ \partial_\mu\alpha^{(0)},\quad
\delta_{\rm s} B^{(n)}_\mu=0, \quad
\delta_{\rm s} B^{(n)}_5=0\ ,\\
\delta_{\rm s}\Phi^{(0)} & =\ -\left( ig\frac{\sigma^i}{2}\alpha^{(0)i}+ig'\frac{Y}{2}\alpha^{(0)} \right)\Phi^{(0)},\quad\delta_{\rm s}\Phi^{(n)}=-\left( ig\frac{\sigma^i}{2}\alpha^{(0)i}+ig'\frac{Y}{2}\alpha^{(0)} \right)\Phi^{(n)},
\end{align}
where the $\alpha^{(0)i}$ and $\alpha^{(0)}$ represent, respectively, the KK zero modes of the $SU_5(2)_W$ and $U_5(1)_Y$ gauge parameters, while $g$ and $g'$ are the dimensionless four--dimensional couplings corresponding, respectively, to the $SU_4(2)_{\rm L}$ and $U_4(2)_Y$ gauge groups. The relations linking the four--dimensional couplings with the extra--dimensional ones are $g\sqrt{2\pi R}=g_5$ and $g'\sqrt{2\pi R}=g_5'$. We remark that this set of gauge transformations are defined exclusively by zero modes of gauge parameters. Note also that the zero modes $W^{(0)i}_\mu$, $B^{(0)}_\mu$ transform as gauge fields, while the corresponding KK excited modes $W^{(n)i}_\mu$ and $B^{(n)}_\mu$, as well as the scalars $W^{(n)i}_5$ and  $B^{(n)}_5$, transform as matter fields. There is an infinite number of other gauge transformations, the NSGT, that possess an involved structure
\begin{align}
\delta_{\rm ns} W^{(0)i}_\mu & = g\epsilon^{ijk}W^{(n)j}_\mu\alpha^{(n)k},\quad\delta_{\rm ns} W^{(n)i}_\mu={\cal D}^{(nm)ij}_\mu\alpha^{(m)j},\quad\delta_{\rm ns} W^{(n)i}_5={\cal D}^{(nm)ij}_5\alpha^{(m)j}, \\
\delta_{\rm ns} B^{(0)}_\mu & =0,\quad\delta_{\rm ns} B^{(n)}_\mu=\partial_\mu\alpha^{(n)},\quad\delta_{\rm ns} B^{(n)}_5=-\frac{n}{R}\alpha^{(n)}\ , \\
\delta_{\rm ns}\Phi^{(0)} & =-\left( ig\frac{\sigma^i}{2}\alpha^{(n)i}+ig'\frac{Y}{2}\alpha^{(0)} \right)\Phi^{(n)},\quad\delta_{\rm ns}\Phi^{(n)}=-\left( ig\frac{\sigma^i}{2}\alpha^{(m)i}+ig'\frac{Y}{2}\alpha^{(m)} \right)\left( \delta^{nm}\Phi^{(0)}+\Delta^{nsm}\Phi^{(s)} \right),
\end{align}
where we have defined
\begin{eqnarray}
{\cal D}^{(nm)ij}_\mu&=&\delta^{nm}{\cal D}^{(0)ij}_\mu-g\epsilon^{ijk}\Delta^{nsm}W^{(s)k}_\mu,
\\ \nonumber \\
{\cal D}^{(nm)ij}_5&=&-\delta^{nm}\delta^{ij}\frac{n}{R}-g\epsilon^{ijk}\Delta'^{nsm}W^{(s)k}_5.
\end{eqnarray}
The symbol $\Delta^{nsm}$ is defined in terms of products and sums of Kronecker deltas. However, the explicit form of this object is not necessary to achieve the goals of this paper. Einstein's summation convention is also used for Fourier modes, each sum starting from 1. The NSGT are determined just by excited modes of gauge parameters, in this case by $\alpha^{(n)j}$ and $\alpha^{(n)}$ which correspond to excited modes of  $SU_5(2)_W$  and $U_5(1)_Y$ gauge paramters. The form of these transformations reveal a quite different nature of the KK excited modes $W^{(n)j}_\mu$ when comparing them with the corresponding SGT; these fields transform as gauge fields through the object ${\cal D}^{(mn)ij}_\mu$. The excited modes $B^{(n)}_\mu$ are clearly also gauge fields with respect to the NSGT. The zero modes $W^{(0)}_\mu$, on the other hand, have a transformation law under the NSGT that resembles the variations of matter fields, but with a more intricate functional form that contains an infinite sum and indicates that they are not gauge fields under the NSGT. The scalars $W^{(n)i}_5$ and $B^{(n)}_5$ play the the interesting role of pseudo--Goldstone bosons similar to those occurring  in the Higgs mechanism. However, in contrast with systems where spontaneous symmetry breaking takes place, in the compactification scenario there are not broken gauge generators~\cite{LMNT}. There exists a gauge in which these type of pseudo--Goldstone bosons are removed from the effective theory, and simultaneously   the KK excited modes of gauge fields with four--dimensional spacetime indices become massive~\cite{NT1}. This fact is remarkable and indicates that some KK excited modes can be turn massive, no matter whether they come from a massless five--dimensional field.

\subsection{A covariant gauge--fixing procedure}
The divergent behaviour of path integrals involved in the quantization of gauge systems arises because of the existence of a set of physically equivalent configurations connected to each other by gauge transformations. The inclusion in path integrals of extra degrees of freedom due to gauge invariance triggers such divergent comportment, which must be removed by fixing the gauge. In the literature~\cite{edgf}, different schemes to fix the gauge in extra--dimensional gauge theories have been proposed. In the present paper, we use the gauge--fixing procedure, of renormalizable type, that was propounded in Refs.~\cite{NT1,CGNT}.  This method is based on the covariant gauge--fixing scheme given in Ref.~\cite{MTTR}, where it was applied to the so--called 331 models~\cite{331}; it shows an unconventional approach possessing the spirit of other schemes such as the one introduced by Fujikawa~\cite{Fuji}, the background--field method~\cite{bfm}, and the pinch technique~\cite{pt}. The main feature of these approaches is that they yield quantized theories in which part of the gauge invariance still remains. Since the KK theory discussed in this paper is separately invariant under the SGT and NSGT, it gives us the possibility to remove the gauge invariance part associated to the NSGT, and maintain SGT invariance of the quantum Lagrangian. The approach followed in Refs.~\cite{NT1,CGNT} to fix the gauge in extra--dimensional gauge theories not only is interesting from a theoretical perspective, but also implies valuable simplifications in phenomenological calculations~\cite{FMNRT}. This useful behavior has been also pointed out and exploited in contexts other than extra--dimensional theories~\cite{MTTR}.

A complete discussion on the quantization of the KK theory conceived in the 5DSM can be found in Ref.~\cite{CGNT}, where the derivation of the quantum Lagrangian, ${\cal L}_{\rm KK}^{\rm q}$, is carried out within the framework of Becchi--Rouet--Stora--Tyutin symmetry~\cite{BRST}, and the result is expressed as
\begin{equation}
{\cal L}_{\rm KK}^{\rm q}={\cal L}_{\rm KK}+{\cal L}_{\rm GF}+{\cal L}_{\rm FPG} ;
\end{equation}
where ${\cal L}_{\rm GF}$ represents the gauge--fixing term and ${\cal L}_{\rm FPG}$ stands for the Faddeev--Popov ghost term. The gauge fixing term is given by
\begin{equation}
{\cal L}_{\rm GF}=-\frac{1}{2\xi}\left( f^{(n)i}_\xi f^{(n)i}_\xi+f^{(n)}_\xi f^{(n)}_\xi \right),
\end{equation}
with gauge--fixing functions $ f^{(n)i}_\xi $ defined as
\begin{equation}
f^{(n)i}_\xi={\cal D}^{(0)ij}_\mu W^{(n)j\mu}-\xi\left( \frac{n}{R} \right)W^{(n)i}_5+ig\xi\left( \Phi^{(n)\dag}\frac{\sigma^i}{2}\Phi^{(0)}-\Phi^{(0)\dag}\frac{\sigma^i}{2}\Phi^{(n)} \right)
\end{equation}
introduced in the $W^{(n)i}_\mu$ sector, and gauge--fixing functions $ f^{(n)}_\xi $
\begin{equation}
f^{(n)}_\xi=\partial_\mu B^{(n)\mu}-\xi\left( \frac{n}{R} \right)B^{(n)}_5+ig'\xi\left( \Phi^{(n)\dag}\frac{Y}{2}\Phi^{(0)}-\Phi^{(0)\dag}\frac{Y}{2}\Phi^{(n)} \right)
\end{equation}
supplied for the $B^{(n)}_\mu$ sector. These functions, which transform covariantly under the SGT, incorporate gauge dependence through the gauge--fixing parameter, denoted by $\xi$. It is worth to mention that these gauge functions introduce modifications to some vertices of the theory, among which one finds the elimination of the unphysical bilinear and trilinear couplings $W^{(n)i}_\mu W^{(n)i}_5$ and $W^{(0)i}_\mu W^{(n)j\mu} W^{(n)k}_5$, as well as the couplings $B^{(n)}_\mu B^{(n)}_5$. The gauge--fixing functions also enter the Faddeev--Popov ghost sector, which then inherits gauge dependence. This covariant gauge--fixing procedure introduces simplifications on phenomenological calculations that involve pseudo--Goldstone bosons and ghost fields; to be more explicit, the contributions from the ghost sector are equal to minus twice the contributions produced by the pseudo--Goldstone bosons.

\subsection{Definitions of mass eigenstates}
In this section, we provide the definitions of mass eigenstates of KK excited modes. The mass eigenstates of SM fields are defined as usual. Recall that at tree level all masses of KK excited modes come from both, compactification and spontaneous symmetry breaking. Through the rest of this paper, we will use the definition $m_n=n/R\equiv n M_{\rm c}$, where $M_{\rm c}$ represents the compactification scale.

The mass of a KK zero--mode field $\varphi^{(0)}$, which is a SM dynamical variable, is denoted by $m_{\varphi^{(0)}}$.  The mass eigenstates for the KK excited gauge modes $W^{(n)i}_\mu$ and $B^{(n)}_\mu$ are defined, in an analogous way to the SM fields, as
\begin{eqnarray}
W^{(n)+}_\mu&=&\frac{1}{\sqrt{2}}\left( W^{(n)1}_\mu-iW^{(n)2}_\mu \right),
\\ \nonumber \\
W^{(n)-}_\mu&=&\frac{1}{\sqrt{2}}\left( W^{(n)1}_\mu+iW^{(n)2}_\mu \right),
\\ \nonumber \\
Z^{(n)}_\mu&=&c_{\rm w}W^{(n)3}_\mu-s_{\rm w}B^{(n)}_\mu,
\\ \nonumber \\
A^{(n)}_\mu&=&s_{\rm w}W^{(n)3}_\mu+c_{\rm w}B^{(n)}_\mu,
\end{eqnarray}
where $s_{\rm w}$ and $c_{\rm w}$ denote the sine and cosine of the weak mixing angle, respectively. Notice that the we make use of the same rotation matrix that defines the SM photon and the $Z$ boson to define their KK--excited--modes replicas. Masses of these KK excited modes are
\begin{align}
m_{W^{(n)}} & =\sqrt{m_{W^{(0)}}^2+m_n^2} \quad\textrm{for}\quad W^{(n)\pm}_\mu\ , \\
m_{Z^{(n)}} & =\sqrt{m_{Z^{(0)}}^2+m_n^2}\quad\textrm{for}\quad Z^{(n)}_\mu\ ,  \\
m_{A^{(n)}} & =m_n \quad\textrm{for}\quad  A^{(n)}_\mu\ .
\end{align}

The KK excited modes of the Higgs doublet are arranged as
\begin{equation}
\Phi^{(n)}=
\left(
\begin{array}{c}
\phi^{(n)+}
\\ \\
H^{(n)}+i\phi_{\rm im}^{(n)}
\end{array}
\right)
\end{equation}
where $H^{(n)}$ and $\phi^{(n)}_{\rm im}$ are real scalar fields, whereas $\phi^{(n)+}$ is a charged field that is related to the non--self--conjugate field $\phi^{(n)-}$ by $\phi^{(n)-}=\left(\phi^{(n)+}\right)^{\dag}$. The mass of the neutral--scalar $H^{(n)}$ is related to that of the SM Higgs boson, so that it is given by
\begin{equation}
m_{H^{(n)}}=\sqrt{m_{H^{(0)}}^2+m_n^2}\ .
\end{equation}
We define now the charged unphysical scalars $W^{(n)\pm}_5=\frac{1}{\sqrt{2}}\left( W^{(n)1}_5\mp iW^{(n)2}_5 \right)$, and they will be combined with the charged scalars $\phi^{(n)\pm}$ to define the charged physical scalars $H^{(n)\pm}$, with mass $m_{W^{(n)}}$, and the pseudo--Goldstone bosons $G^{\pm}_{W^{(n)}}$ as follows:
\begin{eqnarray}
\label{eb1}
H^{(n)+}&=&{\rm cos}(n_{W^{(0)}})\,\phi^{(n)+}+i\,{\rm sin}(n_{W^{(0)}})\,W^{(n)+}_5,
\\ \nonumber \\
\label{eb2}
H^{(n)-}&=&{\rm cos}(n_{W^{(0)}})\,\phi^{(n)-}-i\,{\rm sin}(n_{W^{(0)}})\,W^{(n)-}_5,
\\ \nonumber \\
\label{eb3}
G^+_{W^{(n)}}&=&{\rm sin}(n_{W^{(0)}})\,\phi^{(n)+}-i\,{\rm cos}(n_{W^{(0)}})\,W^{(n)+}_5,
\\ \nonumber \\
\label{eb4}
G^-_{W^{(n)}}&=&{\rm sin}(n_{W^{(0)}})\,\phi^{(n)-}+i\,{\rm cos}(n_{W^{(0)}})\,W^{(n)-}_5,
\end{eqnarray}
 where the mixing angle $n_{W^{(0)}}$ is determined by the relation
\begin{equation}
{\rm tan}\,n_{W^{(0)}}=\frac{m_{W^{(0)}}}{(n/R)}\equiv\frac{m_{W^{(0)}}}{m_n}\ .
\end{equation}

Concerning neutral scalars, we perform the rotation
\begin{equation}
\left(
\begin{array}{c}
W^{(n)3}_5
\\ \\
B^{(n)}_5
\end{array}
\right)=
\left(
\begin{array}{cc}
c_{\rm w}&s_{\rm w}
\\ \\
-s_{\rm w}&c_{\rm w}
\end{array}
\right)
\left(
\begin{array}{c}
\hat{W}^{(n)3}_5
\\ \\
G_{A^{(n)}}
\end{array}
\right),
\end{equation}
which defines the pseudo--Goldstone boson
\begin{equation}
G_{A^{(n)}}=s_{\rm w}\,W^{(n)3}_5+c_{\rm w}\, B^{(n)}_5,
\end{equation}
associated to the gauge boson $A^{(n)}_\mu$. The scalar $\hat{W}^{(n)3}_5$ plays a role in the definition of neutral--boson mass eigenstates as it is merged by a rotation with the neutral scalar $\phi^{(n)}_{\rm im}$ to give rise to
\begin{eqnarray}
h^{(n)}&=&{\rm cos}(n_{Z^{(0)}})\,\phi^{(n)}_{\rm im}-{\rm sin}(n_{Z^{(0)}})\,\hat{W}^{(n)3}_5,
\\ \nonumber \\
G_{Z^{(n)}}&=&{\rm sin}(n_{Z^{(0)}})\,\phi^{(n)}_{\rm im}+{\rm cos}(n_{Z^{(0)}})\,\hat{W}^{(n)3}_5,
\end{eqnarray}
where the mixing angle $n_{Z^{(0)}}$ is obtained from
\begin{equation}
{\rm tan}\,n_{Z^{(0)}}=\frac{m_{Z^{(0)}}}{m_n}.
\end{equation}
The KK--excited field $h^{(n)}$ is a physical neutral scalar\footnote{In Ref.~\cite{CGNT}, the scalar $h^{(n)}$ was denoted by $A^{(n)}$. We have changed this notation in order to avoid confusion involving the KK modes $A^{(n)a}_M$.} with mass $m_{Z^{(n)}}$, whereas $G_{Z^{(n)}}$ is the pseudo--Goldstone boson associated with the KK--excited mode $Z^{(n)}_\mu$.

The definition of fermionic mass eigenstates of KK excited modes is more intricate, as it involves two rotations, the first of which is performed in flavor space. This first step is accomplished similarly to that of the SM fields. In flavor space, one defines
\begin{equation}
{\cal E}^{(n)}_{\rm L,R}=
\left(
\begin{array}{c}
e^{(n)}
\\ \\
\mu^{(n)}
\\ \\
\tau^{(n)}
\end{array}
\right)_{\rm L,R},
\hspace{1cm}
{\cal N}^{(n)}_{\rm L,R}=
\left(
\begin{array}{c}
\nu_e^{(n)}
\\ \\
\nu_\mu^{(n)}
\\ \\
\nu_\tau^{(n)}
\end{array}
\right)_{\rm L,R},
\hspace{1cm}
{\cal \hat{E}}^{(n)}_{\rm L,R}=
\left(
\begin{array}{c}
\hat{e}^{(n)}
\\ \\
\hat{\mu}^{(n)}
\\ \\
\hat{\tau}^{(n)}
\end{array}
\right)_{\rm L,R},
\end{equation}
\begin{equation}
{\cal D}^{(n)}_{\rm L,R}=
\left(
\begin{array}{c}
d^{(n)}
\\ \\
s^{(n)}
\\ \\
b^{(n)}
\end{array}
\right)_{\rm L,R},\hspace{1cm}
{\cal U}^{(n)}_{\rm L,R}=
\left(
\begin{array}{c}
u^{(n)}
\\ \\
c^{(n)}
\\ \\
t^{(n)}
\end{array}
\right)_{\rm L,R},\hspace{1cm}
\hat{\cal D}^{(n)}_{\rm L,R}=
\left(
\begin{array}{c}
\hat{d}^{(n)}
\\ \\
\hat{s}^{(n)}
\\ \\
\hat{b}^{(n)}
\end{array}
\right)_{\rm L,R},\hspace{1cm}
\hat{\cal U}^{(n)}_{\rm L,R}=
\left(
\begin{array}{c}
\hat{u}^{(n)}
\\ \\
\hat{c}^{(n)}
\\ \\
\hat{t}^{(n)}
\end{array}
\right)_{\rm L,R},
\end{equation}
and uses the unitary transformations
\begin{eqnarray}
{\cal E}'^{(n)}_{\rm L,R}=V^e_{\rm L}\,{\cal E}^{(n)}_{\rm L,R},&\hspace{0.5cm}&\hat{\cal E}'^{(n)}_{\rm L,R}=V^e_{\rm R}\,\hat{\cal E}^{(n)}_{\rm L,R},
\\ \nonumber \\
{\cal D}'^{(n)}_{\rm L,R}=V^d_{\rm L}\,{\cal D}^{(n)}_{\rm L,R},&\hspace{0.5cm}&\hat{\cal D}'^{(n)}_{\rm L,R}=V^d_{\rm R}\,\hat{\cal D}^{(n)}_{\rm L,R},
\\ \nonumber \\
{\cal U}'^{(n)}_{\rm L,R}=V^u_{\rm L}\,{\cal U}^{(n)}_{\rm L,R},&\hspace{0.5cm}&\hat{\cal U}'^{(n)}_{\rm L,R}=V^u_{\rm R}\,\hat{\cal U}^{(n)}_{\rm L,R},
\end{eqnarray}
for flavor triplets of charged leptons and quarks. Unitary matrices $V_{\rm L}^e$, $V_{\rm L}^u$ and $V_{\rm L}^d$ are the same we used in order to define the mass eigenstates of fermionic zero modes. Neutrino flavor triplets, ${\cal N}^{(n)}_{\rm L,R}$ are demanded to transform as their corresponding leptonic charged partners, so that
\begin{equation}
{\cal N}'^{(n)}_{\rm L,R}=V^e_{\rm L}\,{\cal N}^{(n)}_{\rm L,R}.
\end{equation}
These transformations allow us to define the mass eigenstates of KK excited modes of neutrinos, which is not the case of charged leptons and quarks, for they must be subjected to a further stage. The idea is that these fermionic fields can be fit into doublets associated to certain KK--flavor space:
\begin{equation}
\left(
\begin{array}{c}
e'^{(n)}
\\ \\
\hat{e}'^{(n)}
\end{array}
\right)_{\rm L,R},\hspace{1cm}\left(
\begin{array}{c}
u'^{(n)}
\\ \\
\hat{u}'^{(n)}
\end{array}
\right)_{\rm L,R},\hspace{1cm}\left(
\begin{array}{c}
d'^{(n)}
\\ \\
\hat{d}'^{(n)}
\end{array}
\right)_{\rm L,R},
\end{equation}
with $e'^{(n)}$ and $\hat{e}'^{(n)}$ collectively denoting charged leptons, $u'^{(n)}$ and $\hat{u}'^{(n)}$ denoting up--type quarks, and $d'^{(n)}$ and $\hat{d}'^{(n)}$ being down--type quarks. The final transformation that completes the deduction of mass eigenstates uses the unitary transformations
\begin{eqnarray}
V_{\rm L}=
\left(
\begin{array}{cc}
\displaystyle{\rm cos}\left( \frac{n_{f^{(0)}}}{2} \right)&\displaystyle{\rm sin}\left( \frac{n_{f^{(0)}}}{2}\right)
\\ \\
\displaystyle{\rm sin}\left( \frac{n_{f^{(0)}}}{2} \right)&\displaystyle-{\rm cos}\left( \frac{n_{f^{(0)}}}{2} \right)
\end{array}
\right),
\hspace{1cm}
V_{\rm R}=
\left(
\begin{array}{cc}
\displaystyle{\rm cos}\left( \frac{n_{f^{(0)}}}{2} \right)&\displaystyle-{\rm sin}\left( \frac{n_{f^{(0)}}}{2}\right)
\\ \\
\displaystyle{\rm sin}\left( \frac{n_{f^{(0)}}}{2} \right)&\displaystyle{\rm cos}\left( \frac{n_{f^{(0)}}}{2} \right)
\end{array}
\right),
\end{eqnarray}
where the mixing angle $n_{f^{(0)}}$ is given by
\begin{equation}
{\rm tan}\,n_{f^{(0)}}=\frac{m_{f^{(0)}}}{m_n}.
\end{equation}
These matrices rotate left and right doublets as
\begin{equation}
\left(
\begin{array}{c}
f'^{(n)}
\\ \\
\hat{f}'^{(n)}
\end{array}
\right)_{\rm L,R}=V_{\rm L,R}\left(
\begin{array}{c}
\tilde{f}^{(n)}
\\ \\
\tilde{\hat{f}}^{(n)}
\end{array}
\right)_{\rm L,R},
\end{equation}
in which the fermionic mass eigenstates are the $\tilde{f}^{(n)}$ and $\tilde{\hat{f}}^{(n)}$ fields. These fields are degenerate as their masses is given by
\begin{equation}
m_{f^{(n)}}=\sqrt{m_{f^{(0)}}^2+m_n^2}.
\end{equation}
Now that we have derived all proper mass eigenstates, we simplify our notation by representing the fermionic mass eigenstates from here on simply by $f^{(n)}$ and $\hat{f}^{(n)}$.

\section{Contributions to the $WWV$ interaction from  universal extra dimensions}
\label{EDWWV}
Through this section, we sketch the calculation of the $CP$--even contributions to the TGC's $WWV$, with $V=\gamma^{(0)},Z^{(0)}$, from the SM with one UED. In Ref.~\cite{FMNRT}, a calculation of the extra--dimensional contributions from the gauge sector of the 5DSM to the TGC's $WWV$ was performed. In the present paper we extend that derivation by including contributions from the rest of the extra--dimensional sectors. We first provide such contributions in terms of Passarino--Veltman scalar functions within a heavy--compactification scenario in which the compactification scale, $M_{\rm c}=1/R$, is large enough to take the mass spectrum of the KK excited modes to be degenerate. In other words, we assume that the mass $m_{\varphi^{(n)}}$ of any KK excited mode $\varphi^{(n)}$ is given by $m_{\varphi^{(n)}}\approx m_n$. Then, under such circumstances, we solve the scalar functions and express the extra--dimensional contributions in terms of elementary functions. The whole calculation is made in the covariant gauge--fixing procedure of Ref.~\cite{CGNT} withinn the non--linear Feynman--'t Hooft gauge, $\xi=1$. We take the $W$ bosons on--shell but keep the neutral $V$ bosons off--shell.

\subsection{Parametrization of new--physics effects on the $WWV$ vertex}
In general, the search for new--physics can be carried out either  by directly producing new heavy states at colliders or by investigating the heavy--physics virtual effects on SM observables. Even if the production of KK modes could be achieved at the Large Hadron Collider (LHC), their determination would be challenging~\cite{CMS,HP,ilcandlhc}. The cleanest way in which the LHC would determine the extra--dimensional nature of the observed effects is the production of second KK modes. The possibility of the lightest KK particle to model~\cite{ST} dark matter favors a mass scale for such particle as large as $1.3$~TeV~\cite{BKP} , which makes it unlikely to produce these second KK states at the LHC. The usage of a linear collider to study the virtual effects of KK modes on SM observables, on the other hand, would make~\cite{HP,ilcandlhc} the desired discrimination easier to reach, as an unambiguous measure of the spin of a first--stage KK particle could be accomplished, and their interactions with SM particles accurately quantified through the detailed examination of virtual effects of KK modes on low--energy physics. To this respect, the importance of analyses of TGC's lies on the possibility that hints on an ultraviolet completion could manifest as contributions to such interactions. The calculation of corrections from SM extensions to TGC's is an option to study high--energy physics by analyzing the outcomes generated by virtual effects. From the perspective of effective field theory, the effects of physics beyond the SM on TGC's involving charged gauge bosons are expected to be more important than the impact of such new physics on neutral TGC's, for the first new--physics effects on the former are parametrized by nonrenormalizable operators of mass--dimension six, while the less--suppressed contributions to the latter are produced by mass--dimension--eight operators~\cite{LPTT}.

In space--time, the $CP$--even high--energy--physics contributions to the $WWV$ interaction are parametrized~\cite{HPZH} by the Lagrangian
\begin{eqnarray}
\label{Lwwv}
{\cal L}_{WWV}&=&-ig_V\bigg\{ g_1^V\left( W^{(0)+}_{\mu\nu}W^{(0)-\mu}V^{(0)\nu}-W^{(0)-}_{\mu\nu}W^{(0)+\mu}V^{(0)\nu}\right)
\nonumber \\  &&
+\kappa_V\,W^{(0)+}_\mu W^{(0)-}_\nu V^{(0)\mu\nu}
+\frac{\lambda_V}{m_W^2}\,W^{(0)+}_{\mu\nu}W^{(0)-\nu\rho}V^{(0)}_\rho\hspace{0.001cm}^\mu
\bigg\},
\end{eqnarray}
where $W^{(0)\pm}_{\mu\nu}=\partial_\mu W^{(0)\pm}_\nu-\partial_\nu W^{(0)\pm}_\mu$ and $V^{(0)}_{\mu\nu}=\partial_\mu V^{(0)}_\nu-\partial_\nu V^{(0)}_\mu$, while the couplings $g_V$ are given by $g_{\gamma^{(0)}}=s_{\rm w}\,g$ and $g_{Z^{(0)}}=c_{\rm w}\,g$. As we are interested in the contributions from new physics, we are assuming that the parameters in this Lagrangian do not include the SM part, so that they only involve the extra--dimensional contributions. Notice that, as we are parametrizing physics beyond the electroweak scale, the parameter $\lambda_V$ should be proportional to $m_W^2/M_{\rm c}^2$ for a large compactification scale.
\begin{figure}[!ht]
\center
\includegraphics[width=6cm]{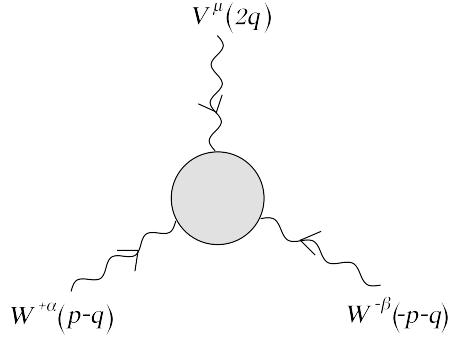}
\caption{\label{wwv} The $WWV$ vertex.}
\end{figure}
Another parametrization, which we will use for our calculations, is given~\cite{wwvp} in the space of vertex functions. Following the conventions of Fig. (\ref{wwv}), the $WWV$ vertex--function parametrization reads
\begin{eqnarray}
\Gamma^V_{\mu\alpha\beta}&=&-ig_V\bigg\{ g_1^V\left[ 2\,p_\mu g_{\alpha\beta}+4(q_\beta g_{\alpha\mu}-q_\alpha g_{\beta\mu}) \right]+2\Delta\kappa_V(q_\beta g_{\alpha\mu}-q_\alpha g_{\beta\mu})
+\frac{4\Delta Q_V}{m_W^2}\,p_\mu\left( q_\alpha q_\beta-\frac{1}{2}q^2g_{\alpha\beta} \right)\bigg\},
\end{eqnarray}
where only the transversal degrees of freedom of the external neutral boson have been left. In the case where the vertex--function parameters are constant, the relations
\begin{eqnarray}
\Delta\kappa_V&=&\kappa_V+\lambda_V-g_1^V,
\\ \nonumber \\
\Delta Q_V&=&-2\lambda_V,
\end{eqnarray}
hold. When taking the external neutral boson to be an on--shell photon, these form factors define the  $CP$--even static electromagnetic properties of the $W$ boson by means of
\begin{eqnarray}
\mu_W&=&\frac{e}{2m_V}(2+\Delta\kappa_\gamma),
\\ \nonumber \\
Q_W&=&-\frac{2}{m_W^2}(1+\Delta\kappa_\gamma+\Delta Q_\gamma),
\end{eqnarray}
where $\mu_W$ is the magnetic dipole moment and $Q_W$ is the electric quadrupole moment.

\subsection{Universal extra dimensions effects on the anomalous $WWV$ form factors}
After performing all KK expansions in the fashion discussed in the last section, integrating out the extra dimension in the effective action, carrying out quantization with the covariant gauge--fixing procedure depicted, and passing to the mass--eigenstates basis, one obtains a KK quantum Lagrangian, ${\cal L}^{\rm q}_{\rm KK}$, that we split as
\begin{equation}
{\cal L}^{\rm q}_{\rm KK}={\cal L}_{\rm SM}+{\cal L}^{\rm 1-loop}_{\rm KK}+{\cal L}^{\rm heavier}_{\rm KK}.
\end{equation}
The ${\cal L}_{\rm SM}$ part represents the low--energy description, that is, the SM, while the term ${\cal L}^{\rm 1-loop}_{\rm KK}$ includes all couplings involving KK zero and excited modes that contribute to SM Green's functions at the one--loop level. The ${\cal L}^{\rm heavier}_{\rm KK}$ Lagrangian contains KK--modes couplings that introduce corrections to low--energy Green's functions at the two--loop level and higher orders. The main objective of the present paper is the calculation of one--loop contributions from the 5DSM to the TGC's $WWV$, so we specifically concentrate on the new--physics effects provided by ${\cal L}^{\rm 1-loop}_{\rm KK}$, which we subdivide into
\begin{equation}
{\cal L}^{\rm 1-loop}_{\rm KK}={\cal L}^{\rm 1-loop}_{\rm G}+{\cal L}^{\rm 1-loop}_{\rm S}+{\cal L}^{\rm 1-loop}_{\rm F}+\cdots,
\end{equation}
where those terms that do not contribute to the one--loop $WWV$ vertex have not been considered and their presence has been only implicitly indicated thorough ellipsis. All the terms shown do produce one--loop corrections to the $WWV$ interaction and have been labeled according to the extra--dimensional sector that gave rise to them: the purely--gauge term ${\cal L}^{\rm 1-loop}_{\rm G}$ involves couplings of gauge bosons, pseudo--Goldstone bosons and ghost fields KK excited modes to gauge zero modes and comes from the five--dimensional gauge sector; the scalar contributing term ${\cal L}^{\rm 1-loop}_{\rm S}$, which is produced by the scalar sector of the 5DSM, comprises tree--level vertices associated to interactions of scalar and gauge KK--excited modes with gauge zero modes, and interactions of scalar KK modes with gauge zero modes as well; and the fermionic sector ${\cal L}^{\rm 1-loop}_{\rm F}$ comes in terms of KK excited fermions coupled to gauge zero modes, whose origin is the fermionic sectors of the extra--dimensional description. The explicit expressions of these Lagrangians can be found in Appendix~\ref{AppL}.

\begin{figure}[!ht]
\center
\includegraphics[width=3.2cm]{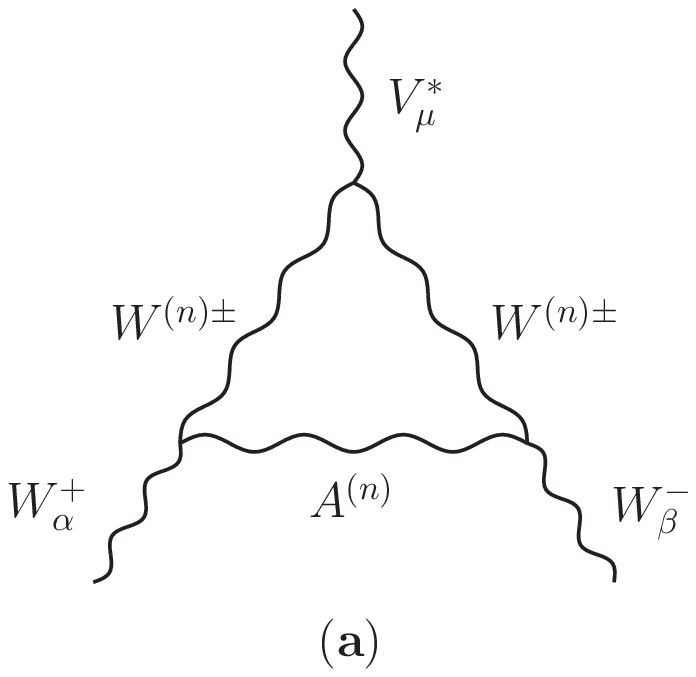}
\hspace{0.5cm}
\includegraphics[width=3.2cm]{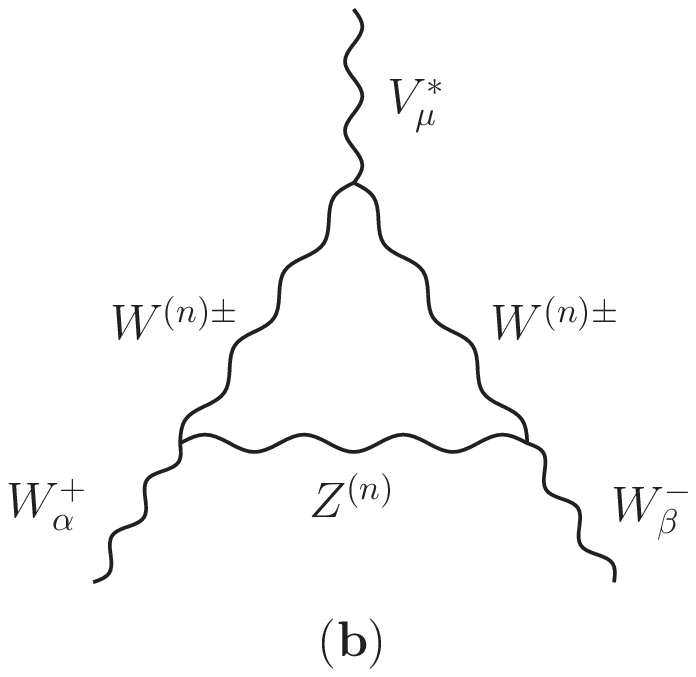}
\hspace{0.5cm}
\includegraphics[width=2.7cm]{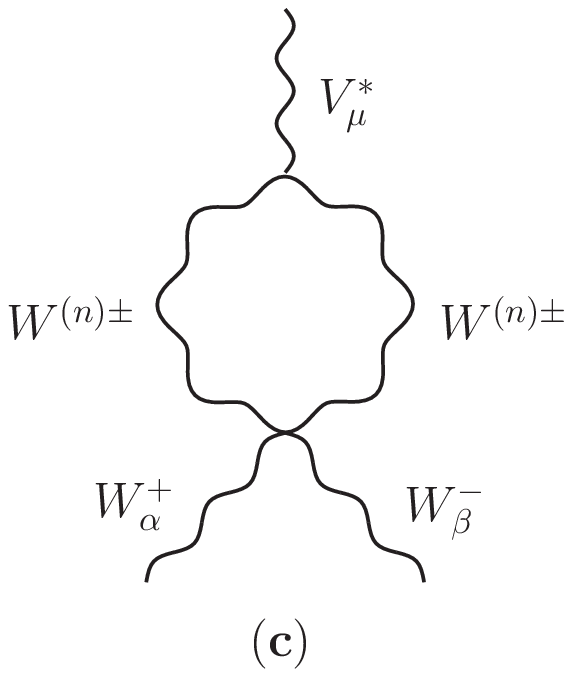}
\\
\vspace{0.5cm}
\includegraphics[width=3.2cm]{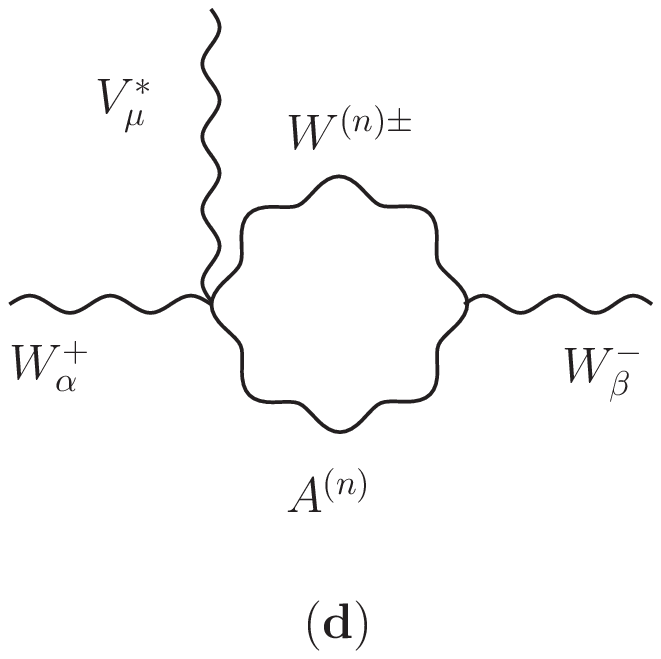}
\hspace{0.5cm}
\includegraphics[width=3.2cm]{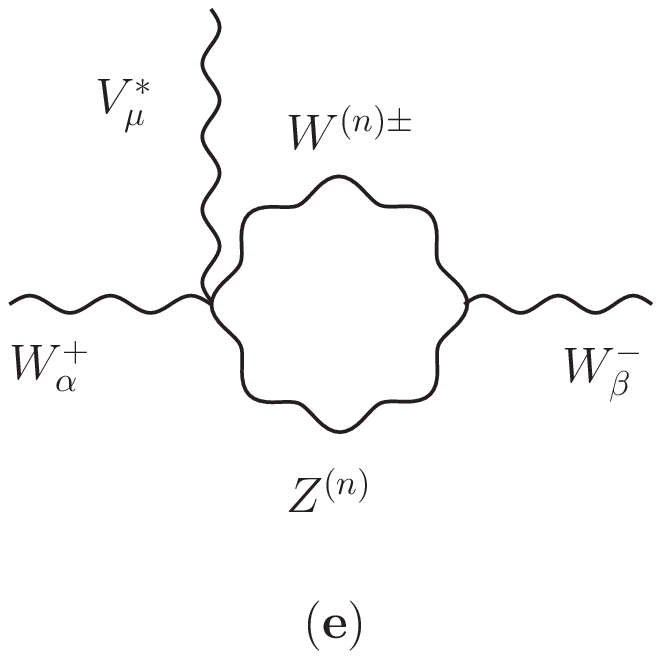}
\hspace{0.5cm}
\includegraphics[width=3.2cm]{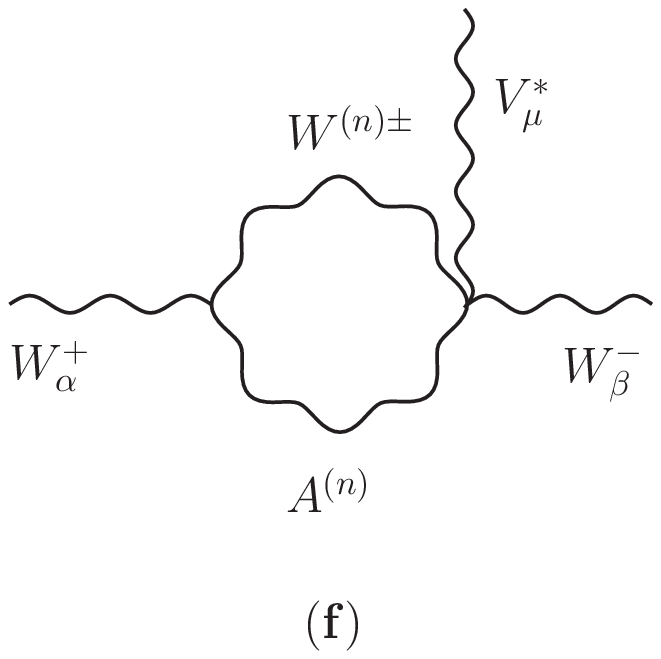}
\\
\vspace{0.5cm}
\includegraphics[width=3.2cm]{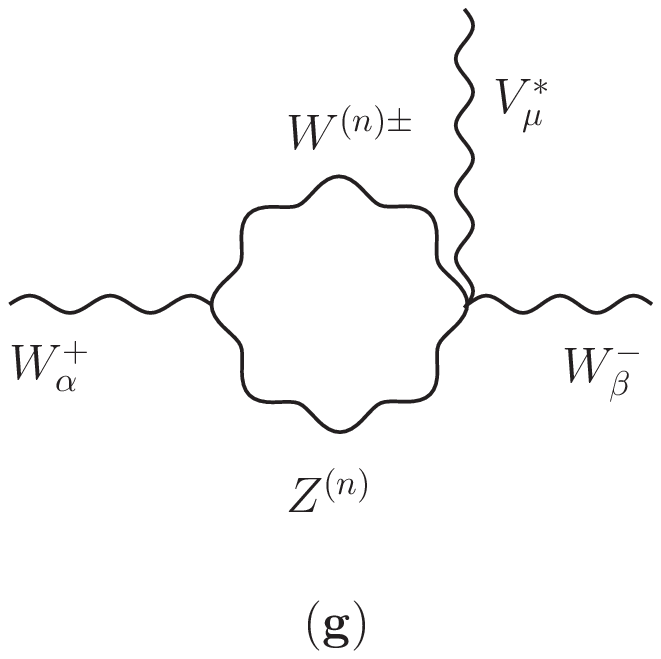}
\hspace{0.5cm}
\includegraphics[width=3.2cm]{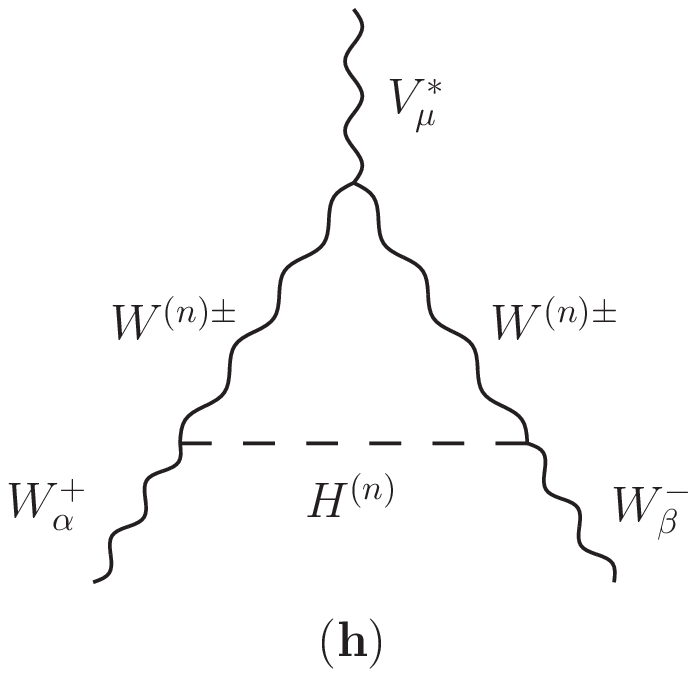}
\hspace{0.5cm}
\includegraphics[width=3.2cm]{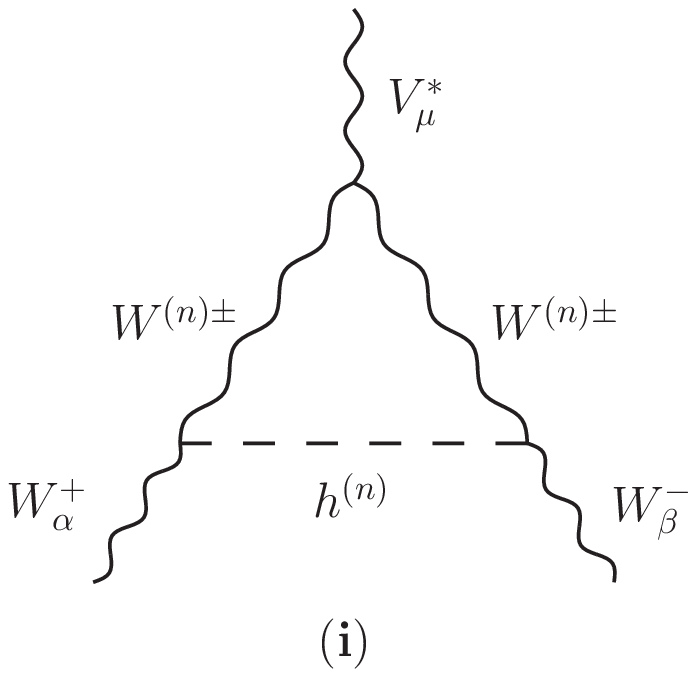}
\caption{\label{gdiag} Contributing one--loop diagrams that involve at least one pure--gauge vertex. Diagrams (a)--(g) only involve KK excited modes that arise from the five--dimensional pure--gauge sector, whereas diagrams (h) and (i) also contain scalar KK excited modes that originate in the five--dimensional Higgs sector.}
\end{figure}

\begin{figure}[!ht]
\center
\includegraphics[width=3.2cm]{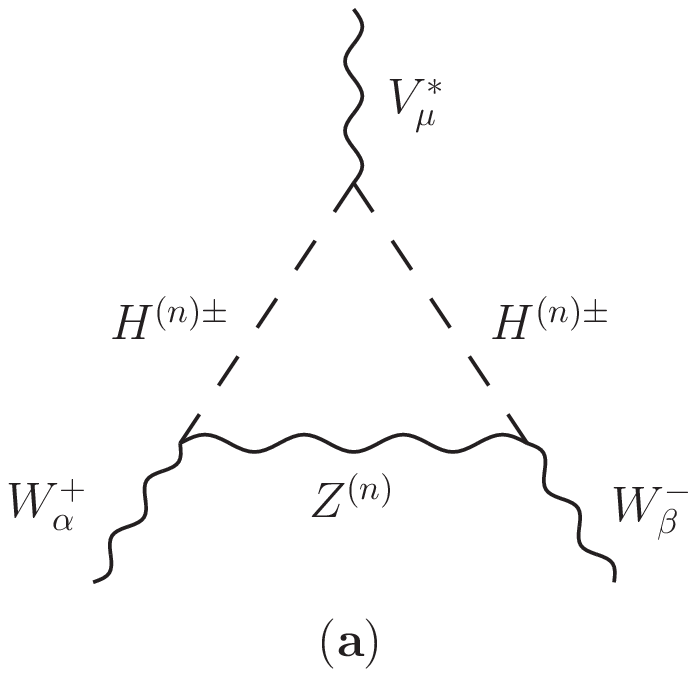}
\hspace{0.5cm}
\includegraphics[width=3.2cm]{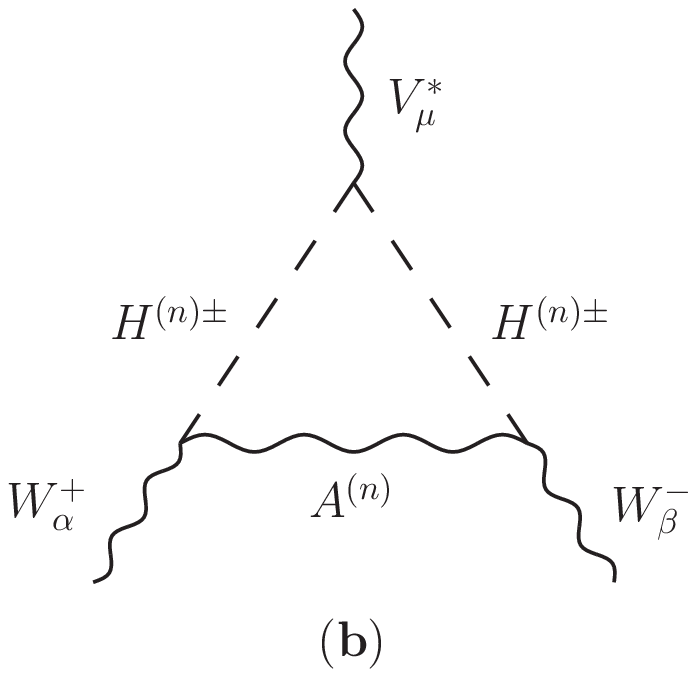}
\caption{\label{gsdiag} One--loop diagrams including an internal gauge KK excited boson and couplings among KK scalar excitations to SM gauge bosons.}
\end{figure}

\begin{figure}[!ht]
\center
\includegraphics[width=3.2cm]{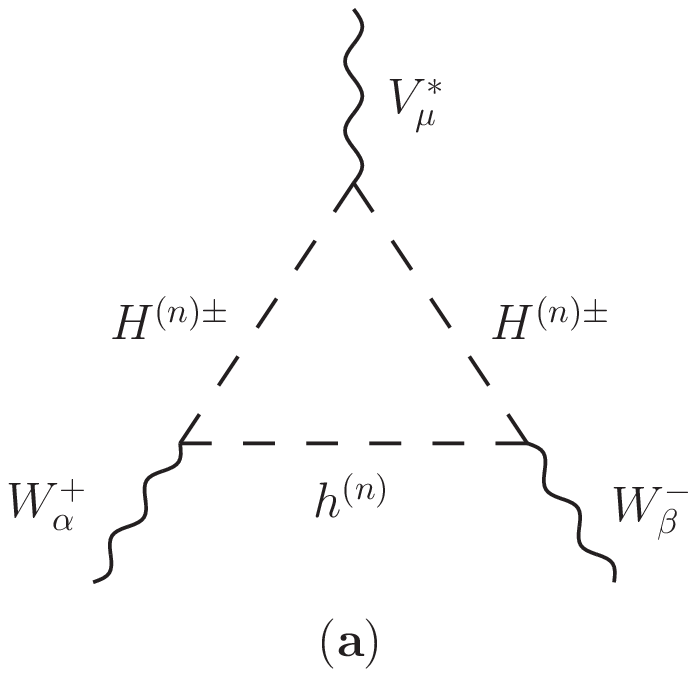}
\hspace{0.5cm}
\includegraphics[width=3.2cm]{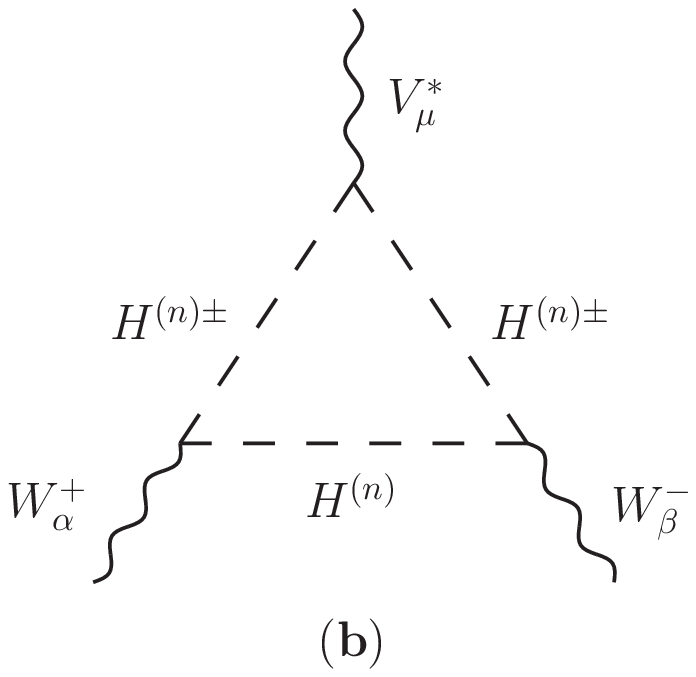}
\hspace{0.5cm}
\includegraphics[width=2.7cm]{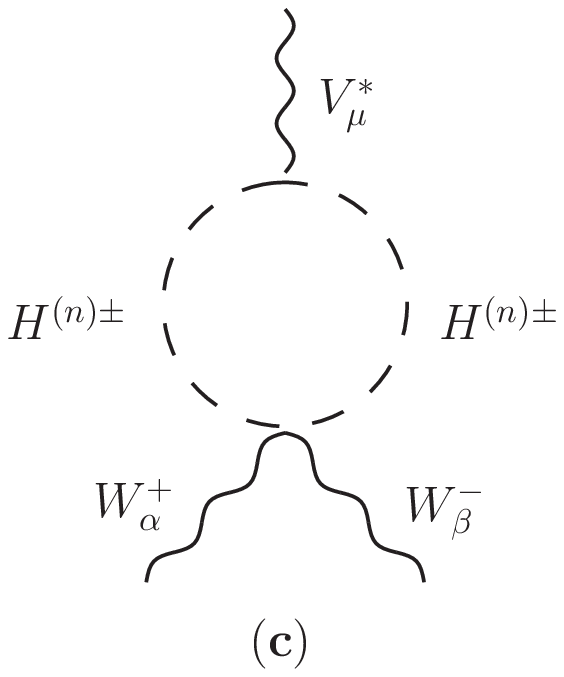}
\hspace{0.5cm}
\includegraphics[width=3.3cm]{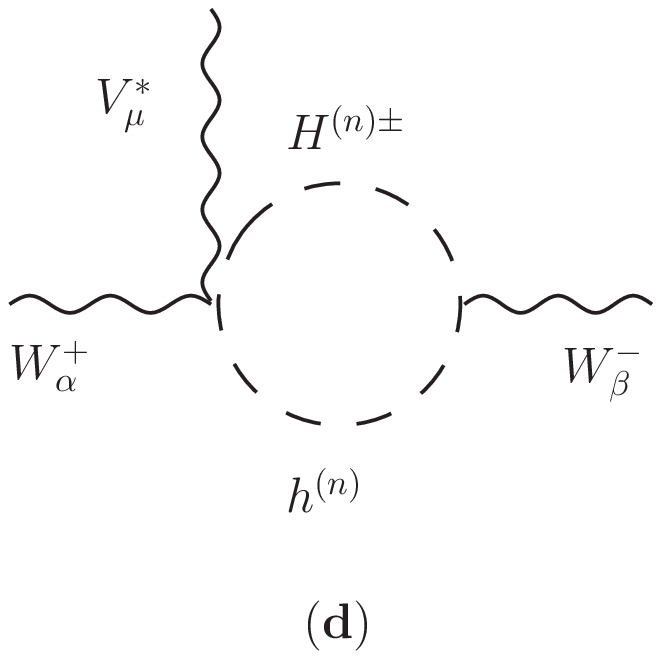}
\hspace{0.5cm}
\includegraphics[width=3.3cm]{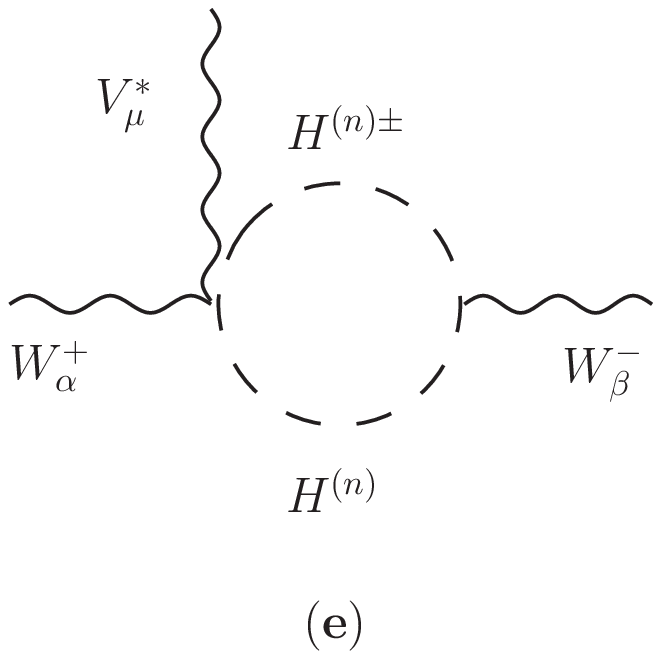}
\hspace{0.5cm}
\includegraphics[width=3.3cm]{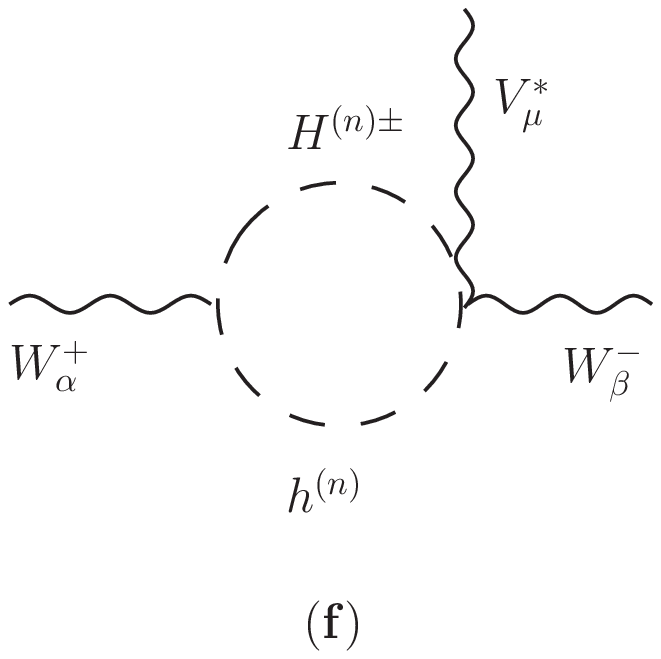}
\hspace{0.5cm}
\includegraphics[width=3.3cm]{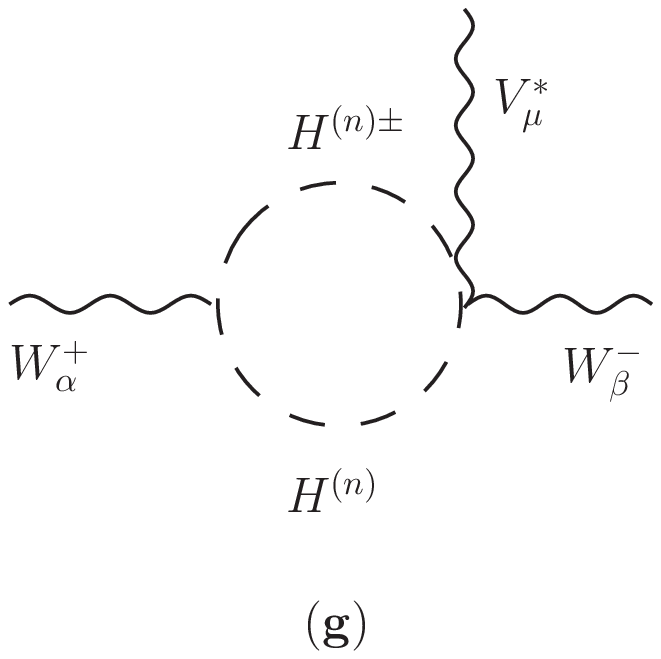}
\caption{\label{sdiag} One--loop diagrams without pure--gauge vertices and involving scalar interactions. The amplitude associated with each diagram  $(c)-(g)$ vanishes identically.}
\end{figure}

\begin{figure}[!ht]
\center
\includegraphics[width=3.2cm]{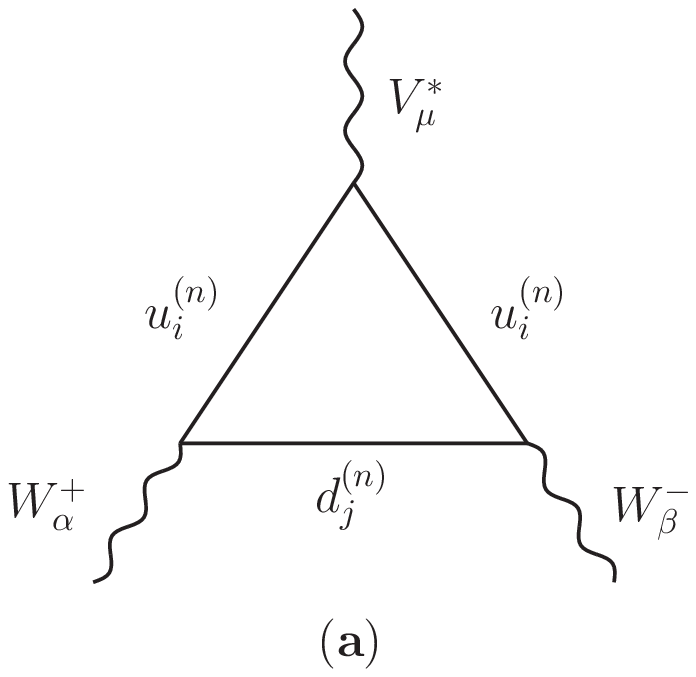}
\hspace{0.5cm}
\includegraphics[width=3.2cm]{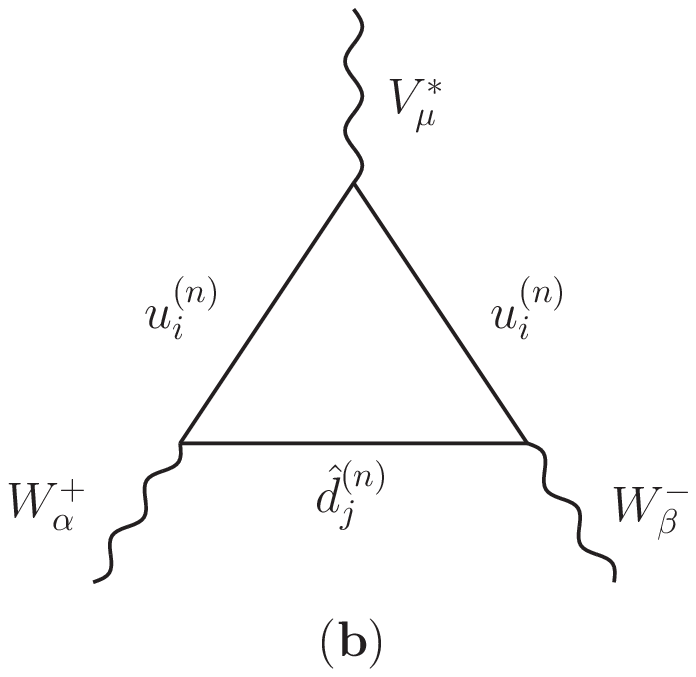}
\hspace{0.5cm}
\includegraphics[width=3.2cm]{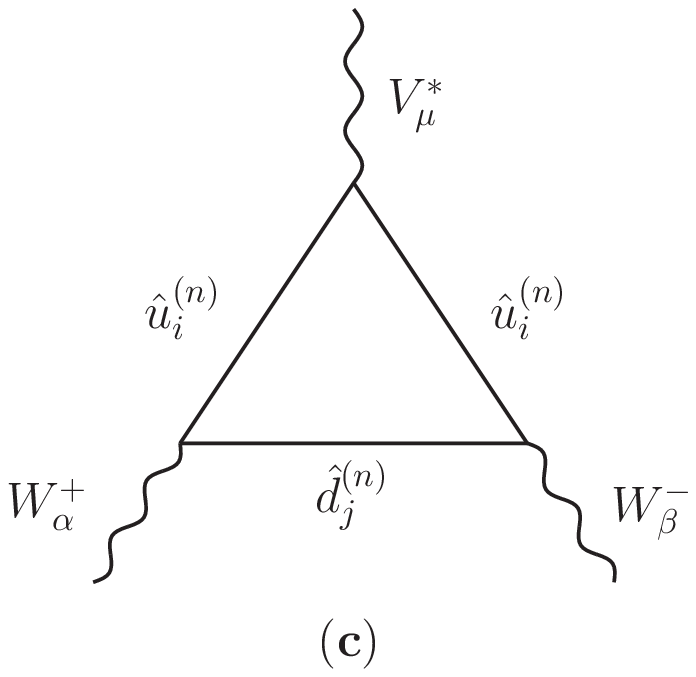}
\hspace{0.5cm}
\includegraphics[width=3.2cm]{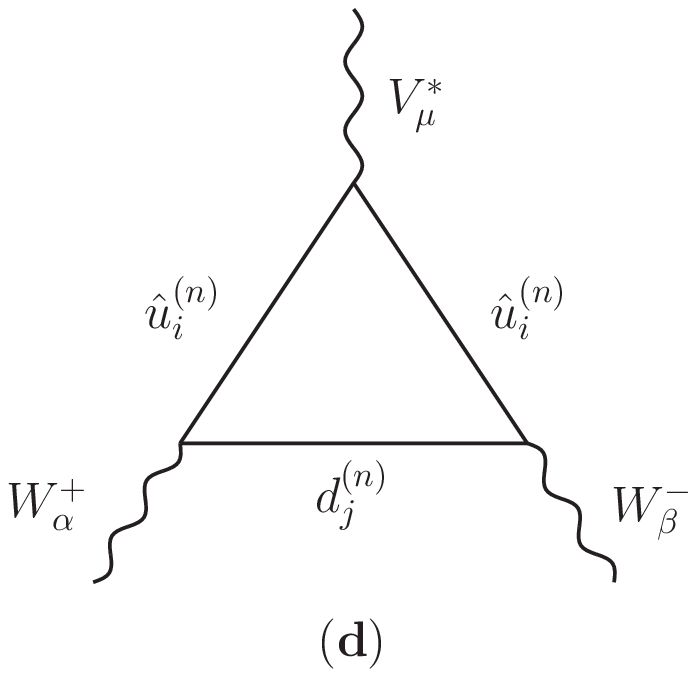}
\\
\vspace{0.5cm}
\includegraphics[width=3.2cm]{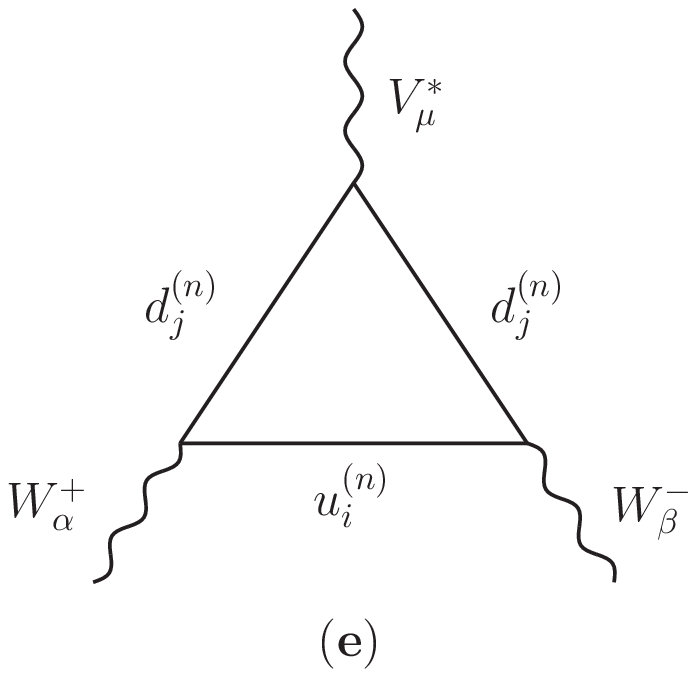}
\hspace{0.5cm}
\includegraphics[width=3.2cm]{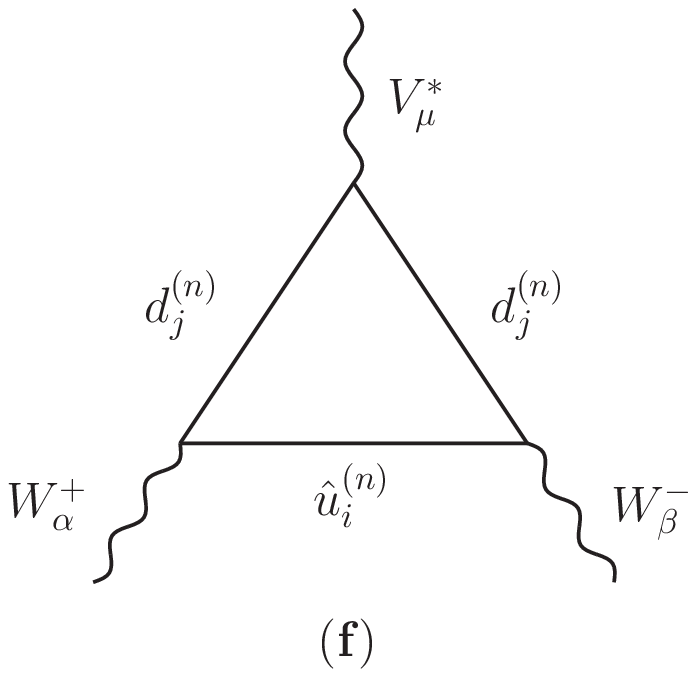}
\hspace{0.5cm}
\includegraphics[width=3.2cm]{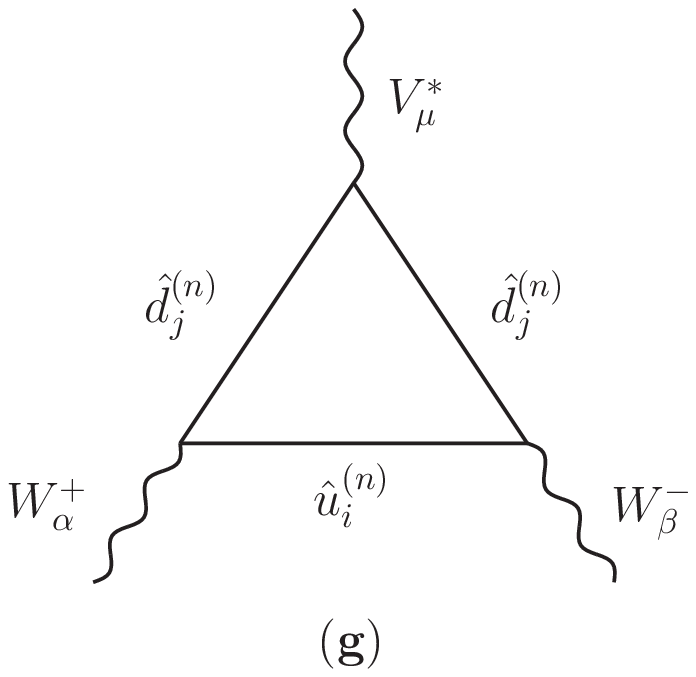}
\hspace{0.5cm}
\includegraphics[width=3.2cm]{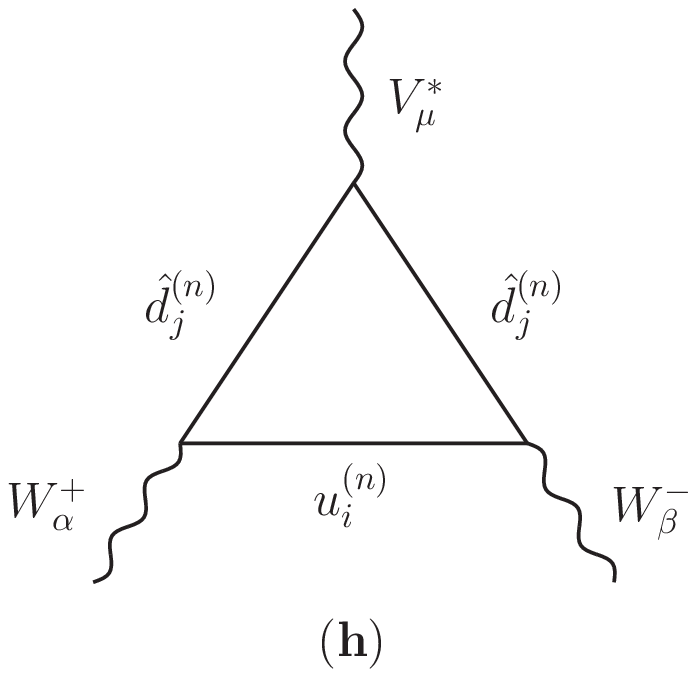}
\\
\vspace{0.5cm}
\includegraphics[width=3.2cm]{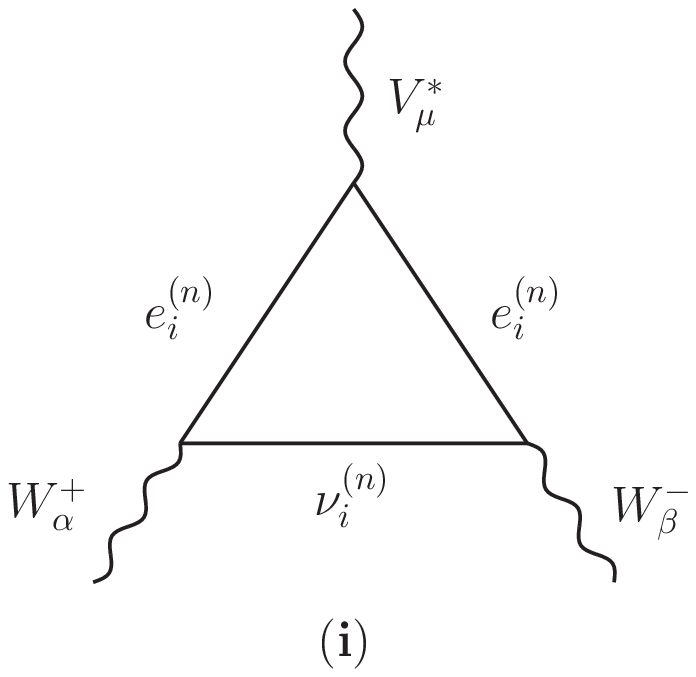}
\hspace{0.5cm}
\includegraphics[width=3.2cm]{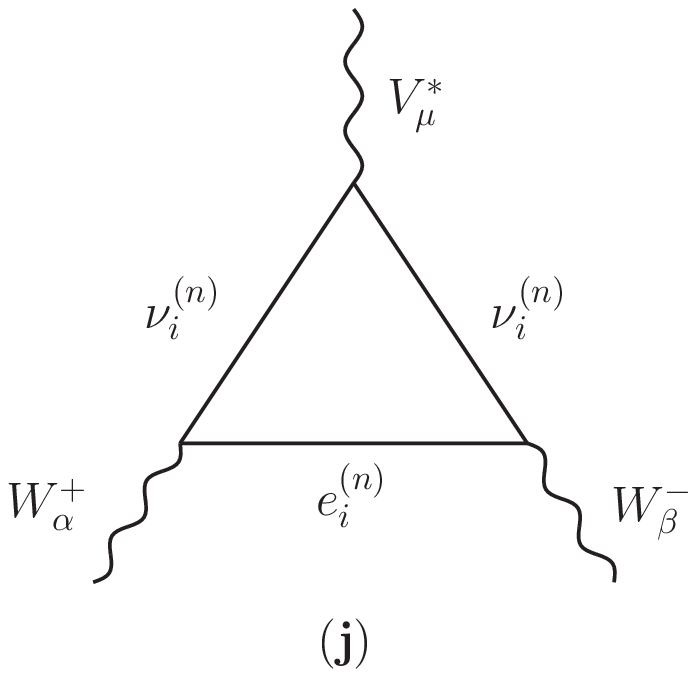}
\hspace{0.5cm}
\includegraphics[width=3.2cm]{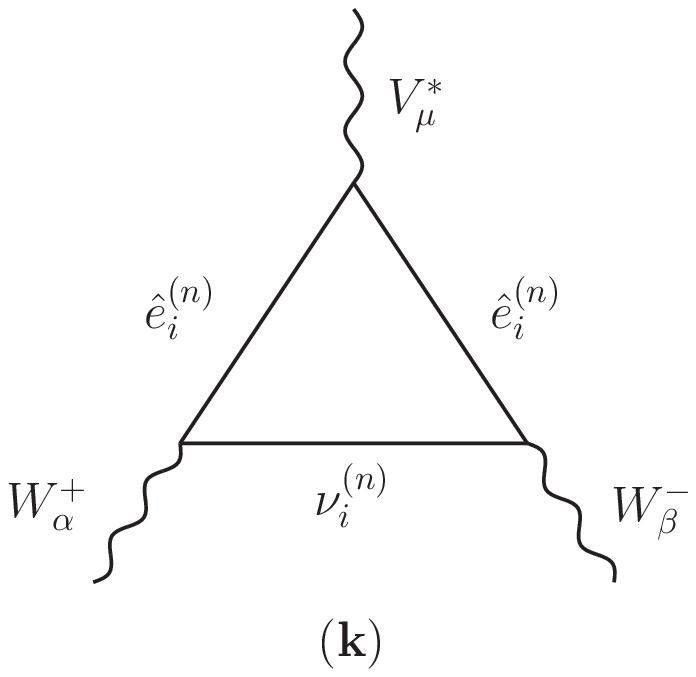}
\hspace{0.5cm}
\includegraphics[width=3.2cm]{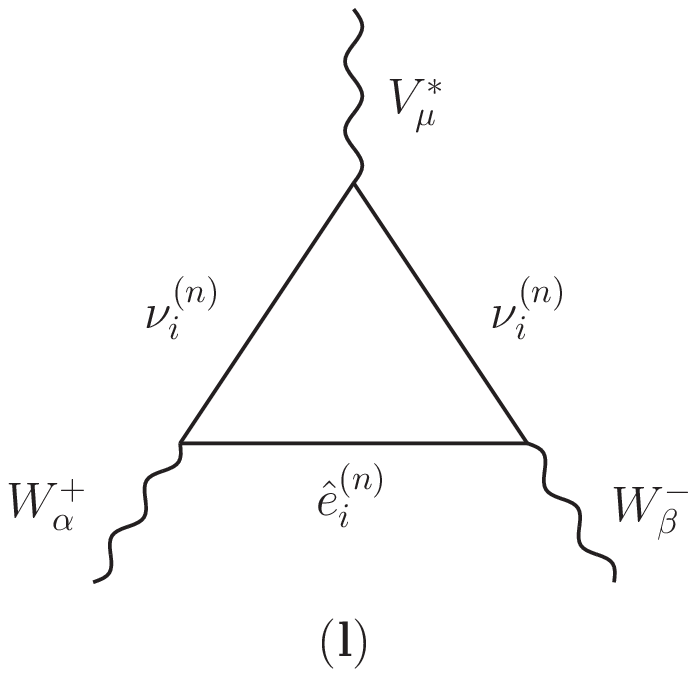}
\caption{\label{fcpediag} One--loop diagrams without pure--gauge vertices and involving fermionic currents. Diagrams (a)--(h) are composed of interactions of KK excited modes of five--dimensional quarks with zero modes of extra--dimensional gauge fields. KK excited modes involved in diagrams (i)--(l) are generated exclusively by five--dimensional leptons.}
\end{figure}

The one--loop extra--dimensional contributions to the $WWV$ vertex are generated by the diagrams shown in Figs. (\ref{gdiag}) to (\ref{fcpediag}). We have found that the extra--dimensional contributions to $g_1^Z$ are, as expected in any renormalizable theory, divergent. The interaction associated to such form factor appears in the SM classical action, so that it must be renormalized. The form factor $\Delta\kappa_V$ is also related to an interaction already present at the classical level, but an anomalous contribution arises in this case. In what follows, we concentrate only on the form factors $\Delta\kappa_V$ and $\Delta Q_V$. We classify the $CP$--even contributions from the 5DSM to such form factors  according to whether the contributing diagrams contain at least one pure--gauge vertex or not. By {\it pure--gauge} vertex we mean a tree--level interaction involving only gauge KK zero and/or excited modes. With this in mind, we express the $\Delta\kappa_V$ and $\Delta Q_V$ form factors as
\begin{eqnarray}
\Delta\kappa_V&=&\Delta\kappa^{\rm GNBC}_V+\Delta\kappa^{\rm NGC}_V,
\\ \nonumber \\
\Delta Q_V&=&\Delta Q^{\rm GNBC}_V+\Delta Q^{\rm NGC}_V.
\end{eqnarray}
The diagrams contributing to $\Delta\kappa_V^{\rm GNBC}$ and $\Delta Q_V^{\rm GNBC}$, which we exhibit in Fig.~(\ref{gdiag}), incorporate one or more pure--gauge vertices. In the case of such diagrams, the external SM neutral boson always couples to a pair of gauge KK excited modes, although the SM $W$ bosons can couple to a scalar, as it occurs in diagrams (h) and (i). On the other hand, the structure of any pure--gauge coupling involving SM bosons and KK excited modes is not distorted by spontaneous symmetry breaking, which is a situation that particularly occurs in the case in which a neutral SM boson is involved. This suggests that the contributions from these diagrams are governed by the $SU_4(2)_{\rm L}$ gauge group, so that we refer to them as {\it gauge neutral--boson contributions} (GNBC). For the $\Delta\kappa_V^{\rm NGC}$ and $\Delta Q_V^{\rm NGC}$ form factors, whose contributing diagrams have been distributed in Figs.~(\ref{gsdiag}) to (\ref{fcpediag}), we use the name {\it non--gauge contributions} (NGC), because, contrastingly to the GNBC, they are affected by electroweak symmetry breaking. The diagrams associated to NGC come from the five--dimensional scalar sector and from the fermionic one, being different to each other in the way that KK excited modes couple to SM gauge bosons. We have separated the NGC into: $i$) those contributions, denoted by $\Delta\kappa_V^{\rm GS}$ and $\Delta Q_V^{\rm GS}$, generated by diagrams that include three--level couplings of a SM neutral boson to scalars and an internal line representing a gauge KK excited mode, which we exhibit in Fig.~(\ref{gsdiag}); $ii$) those coming from diagrams incorporating scalar loops, shown in Fig.~(\ref{sdiag}), and referred to as $\Delta\kappa_V^{\rm S}$ and $\Delta Q_V^{\rm S}$; and $iii$) those represented by $\Delta\kappa_V^{\rm F}$ and $\Delta Q_V^{\rm F}$ and produced by diagrams involving fermionic loops, given in Fig.~(\ref{fcpediag}). This is written as
\begin{eqnarray}
\Delta\kappa_V^{\rm NGC}&=&\Delta\kappa_V^{\rm GS}+\Delta\kappa_V^{\rm S}+\Delta\kappa_V^{\rm F},
\\ \nonumber \\
\Delta Q_V^{\rm NGC}&=&\Delta Q_V^{\rm GS}+\Delta Q_V^{\rm S}+\Delta Q_V^{\rm F}.
\end{eqnarray}
A distinctive feature of the UED framework, first pointed out in Ref.~\cite{ACD}, is the particular importance of one--loop calculations in models incorporating this sort of extra dimensions. The source of this meaningful property is conservation of momentum in compact extra dimensions, which, at the four--dimensional level, is translated into KK--number conservation. The $S^1/Z_2$ orbifold compactification  breaks conservation of KK number to conservation of KK parity, which has the consequence that KK excited modes cannot be single--produced at tree level. This means that there are no tree--level contributions to low--energy observables, but the very first corrections enter at the one--loop level. Indeed, this is the reason behind the relatively weak constraints on the compactification scale associated to this sort of extra--dimensional models. The ATLAS Collaboration performed~\cite{SUSYdp} an analysis that centered in searching for squarks and gluinos in events containing jets, missing transverse momentum and no electrons or muons. The results of such paper were reinterpreted~\cite{ATLASued} in the context of UED, setting the bound $M_{\rm c}>600\,{\rm GeV}$ under sensible assumptions related to the energy scale associated with the higher--energy description lying beyond extra--dimensional physics. A recent investigation~\cite{BBBKP}, which combined the latest LHC data provided by the ATLAS and CMS collaborations, constrained the compactification scale, finding that $M_{\rm c}>500$\,GeV is still allowed within a very narrow region around a value for the mass of the Higgs boson amounting to $m_{H^{(0)}}=125$\,GeV, while around $m_{H^{(0)}}=118$\,GeV only a compactification scale as large as $M_{\rm c}>1000$\,GeV would be still consistent. We then consider the heavy--compactification scenario, which we define by the relation $Q^2<<M_{\rm c}^2$, with $Q=2q$ standing for the incoming momentum of the neutral gauge boson $V$. This framework allows us to take the compactification scale to be large enough so that the mass spectrum of the KK modes is degenerate. In other words, we take the approximation that the mass of any KK excited mode is given by $m_n\approx n/R$. All the results that we show below are given in terms of the dimensionless variables $y_\varphi\equiv m_{\varphi^{(0)}}^2/m_n^2$ and $y_Q=Q^2/m_n^2$. The degenerate mass spectrum of KK modes greatly simplifies all results, which can be written in terms of only four Passarino--Veltman scalar functions. In order not to clutter up the notation, we define $B_0(1)\equiv B_0(0,m_n^2,m_n^2)$, $B_0(2)\equiv B_0(m_{W^{(0)}}^2,m_n^2,m_n^2)$, $B_0(3)\equiv B_0(Q^2,m_n^2,m_n^2)$, and $C_0(m_{W^{(0)}}^2,m_{W^{(0)}}^2,Q^2,m_n^2,m_n^2,m_n^2)$. All results are shown in such a way that the cancelation of ultraviolet divergences is explicit.

As we noticed, the GNBC are supplied by the one--loop diagrams shown in Fig. (\ref{gdiag}). It is important to stress that, in addition to these diagrams, the contributions from pseudo--Goldstone bosons and ghost fields are taken into account, even when the corresponding diagrams are not shown in Fig. (\ref{gdiag}). The diagrams incorporating pseudo--Goldstone bosons and ghosts are similar to the diagrams solely containing gauge--bosons, and all of them, but the triangular ones, vanish. The expressions for the GNBC read
\begin{eqnarray}
\Delta\kappa_V^{\rm GNBC}&=&\frac{g^2}{96\pi^2}\sum_{n=1}^\infty
\frac{1}{\left(4 y_W-y_Q\right){}^3}\Bigg\{640 y_Q y_W \left[B_0(1)-2B_0(2)+B_0(3)\right]-1280 y_W^2
\left[B_0(1)-3B_0(2)+2B_0(3)\right]
\nonumber \\ \nonumber&&
-80y_Q^2\left[B_0(1)-B_0(2)\right]+\frac{1}{y_W+1}\Big(-96
   y_Q^3 \left(y_W+1\right)+2 y_Q^2 y_W \left(554
   y_W+581\right)
\\ \nonumber &&
   -8 y_Q y_W^2 \left(617
   y_W+671\right)+288 y_W^3 \left(20
   y_W+23\right)\Big)\left[B_0(2)-B_0(3)\right]
   \\ \nonumber &&
   +\frac{6m_n^2y_W}{y_W+1}\Big(-12 y_Q \left(y_W
   \left(y_W+15\right)+20\right) y_W+3 y_Q^2
   \left(y_W \left(8 y_W+25\right)+20\right)+16
   \left(4 y_W-5\right) y_W^3\Big)C_0
\\  &&
+20
   y_W \left(y_Q-4 y_W\right) \left(3 y_Q+8
   y_W\right)\Bigg\}
\end{eqnarray}
\begin{eqnarray}
\Delta Q^{\rm GNBC}_V&=&\frac{g^2}{96\pi^2}\sum_{n=1}^\infty\frac{1}{\left(4
   y_W-y_Q\right){}^3}\Bigg\{160 y_Q y_W \left[B_0(1)+2B_0(2)-3B_0(3)\right]+\frac{2560y_W^3}{y_Q}\left[B_0(1)-B_0(3)\right]
\nonumber \\ \nonumber &&
-1280 y_W^2 \left[B_0(1)+B_0(2)-2B_0(3)\right]
+\frac{160y_W
   \left(5y_Q y_W^2+y_Q^2 y_W-6y_W^3\right)}{y_Q}\left[B_0(2)-B_0(3)\right]
\\ \nonumber &&
+\frac{240m_n^2y_W \left(-2
y_Q^2 \left(y_W-3\right) y_W+4y_Q\left(y_W-3\right) y_W^2-y_Q^3-4\left(y_W-4\right) y_W^3\right)}{y_Q}C_0
\\  &&
+\frac{40 y_W
   \left(6 y_Q^2 y_W-20 y_Q y_W^2-y_Q^3+48
   y_W^3\right)}{y_Q}\Bigg\}
\end{eqnarray}

The diagrams of Fig.~(\ref{gsdiag}) produce NGC given by
\begin{eqnarray}
\Delta\kappa^{\rm GS}_V&=&\frac{g^2}{96\pi^2c_{\rm w}^2}\sum_{n=1}^\infty
\frac{3G_V\left(2 s_{\rm w}^2-5\right)y_W}{\left(4 y_W-y_Q\right)(1+y_W)} \Big\{ 2\left[B_0(2)-B_0(3)\right]
-m_n^2\left(y_Q-2
   y_W\right)C_0\Big\}
\\ \nonumber \\
\Delta Q_V^{\rm GS}&=&0
\end{eqnarray}
with the definitions
\begin{equation}
G_V=G_{(\gamma^{(0)},Z^{(0)})}=\left( 1,1-\frac{1}{2c_{\rm w}^2}+{\cal O}(y_W) \right).
\end{equation}

The NGC from scalar KK excited modes through diagrams with scalar loops, which we provide in Fig.~(\ref{sdiag}), contribute to the $\Delta\kappa_V$ and $\Delta Q_V$ form factors as
\begin{eqnarray}
\Delta\kappa_V^{\rm S}&=&\frac{g^2}{96\pi^2}\sum_{n=1}^\infty
\frac{G_V\left(4 c_{\rm w}
   \left(c_{\rm w} y_W+1\right)-1\right)}{\left(4
   y_W-y_Q\right){}^3(1+y_W)^2}\Big\{32 y_Q y_W^2\left[B_0(1)-2B_0(2)+B_0(3)\right]
\nonumber \\ \nonumber &&
-64y_W^3\left[B_0(1)-3B_0(2)+2B_0(3)\right]
-4y_Q^2 y_W\left[B_0(1)-B_0(2)\right]
\\ \nonumber &&
-y_W\left(y_Q^2 y_W+26 y_Qy_W^2\right)\left[B_0(2)-B0(3)\right]-6m_n^2y_W\left(-y_Q^2 y_W \left(y_W+3\right)-y_Q
   \left(y_W-12\right) y_W^2\right)C_0
\\  &&
-y_W\left(4 y_Q y_W^2-3 y_Q^2 y_W+32
   y_W^3\right)\Big\}
\end{eqnarray}
\begin{eqnarray}
\Delta Q^{\rm S}_V&=&\frac{g^2}{96\pi^2}\sum_{n=1}^\infty\frac{G_V\left(4 c_{\rm w}\left(c_{\rm w} y_W+1\right)-1\right)}{\left(4
   y_W-y_Q\right){}^3(1+y_W)^2} \Bigg\{-64y_W^3\left[B_0(1)+B_0(2)-2B_0(3)\right]
\nonumber \\ \nonumber &&
+8 y_W^2y_Q \left[B_0(1)+2B_0(2)-3B_0(3)\right] +\frac{128y_W^4 }{y_Q}\left[B_0(1)-B_0(3)\right]
\\ \nonumber &&
-\frac{8y_W^2 \left(-5y_Q y_W^2-y_Q^2 y_W+6y_W^3\right) }{y_Q}\left[B_0(2)-B_0(3)\right]
\\ \nonumber &&
-\frac{12m_n^2y_W^2\left(2y_Q^2 \left(y_W-3\right) y_W-4y_Q \left(y_W-3\right) y_W^2+y_Q^3+4\left(y_W-4\right) y_W^3\right)}{y_Q}C_0
\\ &&
-\frac{2 y_W^2 \left(-6
   y_Q^2 y_W+20 y_Q y_W^2+y_Q^3-48 y_W^3\right)}{y_Q}\Bigg\}
\end{eqnarray}

The NGC from fermionic KK excited modes come from the  diagrams shown in Fig.~(\ref{fcpediag}). Notice that diagrams ({\it k}) and ({\it l} ) involve charged currents characterized by a sort of KK flavor mixing among leptonic KK excited states $e_i^{(n)}$ and $\hat{e}_i^{(n)}$. Each of these couplings incorporates a $\gamma^5$ matrix, but, as there are two of such KK--flavor--changing charged currents in each of these diagrams, the $\gamma^5$ matrices cancel, and the resulting expression is $CP$--even. The NGC from all diagrams to the $\Delta\kappa_V$ and $\Delta Q_V$ form factors are then
\begin{eqnarray}
\Delta\kappa^{\rm F}_V&=&\frac{g^2}{96\pi^2}\sum_{n=1}^\infty\Bigg[
\frac{F_{V,1}}{\left(4 y_W-y_Q\right){}^3} \Big\{-128 y_Q y_W\left[B_0(1)-2B_0(2)+B_0(3)\right]
\nonumber \\ \nonumber &&
+256y_W^2 \left[B_0(1)-3B_0(2)+2B_0(3)\right]+16y_Q^2\left[B_0(1)-B_0(2)\right]
\\ \nonumber &&
+4\left(y_Q^2y_W+26y_Q y_W^2\right)\left[B_0(2)-B_0(3)\right]-24m_n^2\left(y_Q^2 y_W \left(y_W+3\right)+y_Q \left(y_W-12\right) y_W^2\right)C_0
\\ \nonumber &&
-4 \left(-4 y_Q
   y_W^2+3 y_Q^2 y_W-32 y_W^3\right)\Big\}
\\ \nonumber &&
+\frac{F_{V,2}}{\left(4 y_W-y_Q\right){}^3} \Big\{-128 y_Q y_W\left[B_0(1)+4B_0(2)-5B_0(3)\right]
\\ \nonumber &&
+256y_W^2\left[B_0(1)+3B_0(2)-4B_0(3)\right]
+16y_Q^2 \left[B_0(1)+5B_0(2)-6B_0(3)\right]
\\ \nonumber &&
+4\left(y_Q^2y_W+26 y_Qy_W^2\right)\left[B_0(2)-B_0(3)\right]-24m_n^2\left(y^2_Qy_W\left(y_W-5\right)+y_Q^3+y_Q\left(y_W+4\right)y_W^2\right)C_0
\\ &&
-4 \left(-4 y_Q
   y_W^2+3 y_Q^2 y_W-32 y_W^3\right)\Big\}\Bigg]
\end{eqnarray}
\begin{eqnarray}
\Delta Q^{\rm F}_V&=&\frac{g^2}{96\pi^2}\sum_{n=1}^\infty\frac{F_{V,1}+F_{V,2}}{\left(4y_W-y_Q\right){}^3} \bigg\{-32 y_Qy_W\left[B_0(1)+2B_0(2)-3B_0(3)\right]
\nonumber \\ \nonumber &&
+256y_W^2\left[B_0(1)+B_0(2)-2B_0(3)\right]-\frac{512y_W^3}{y_Q}\left[B_0(1)-B_0(3)\right]
\\ \nonumber &&
+\frac{32y_W \left(-5y_Qy_W^2-y_Q^2y_W+6y_W^3\right)}{y_Q}\left[B_0(2)-B_0(3)\right]
\\ \nonumber &&
+\frac{48m_n^2y_W\left(2y_Q^2 \left(y_W-3\right)y_W-4y_Q\left(y_W-3\right) y_W^2+y_Q^3+4\left(y_W-4\right)y_W^3\right)}{y_Q}C_0
\\  &&
+\frac{8 y_W \left(-6 y_Q^2y_W+20 y_Qy_W^2+y_Q^3-48y_W^3\right)}{y_Q}\bigg\},
\end{eqnarray}
with the $F_{V,i}$ overall factors given by
\begin{eqnarray}
F_{\gamma,1}&=&\sum_{\rm families}\left[ |K_{ud}|^2-1+{\cal O}(y_e) \right],
\\ \nonumber \\
F_{\gamma,2}&=&\sum_{\rm families}{\cal O}(y_e),
\\ \nonumber \\
F_{Z,1}&=&\sum_{\rm families}\left[ 1+{\cal O}(y_u,y_d,y_e) \right],
\\ \nonumber \\
F_{Z,2}&=&\sum_{\rm families}{\cal O}(y_e).
\end{eqnarray}
In performing this calculation, we have neglected the contributions involving off--diagonal entries of the Kobayashi--Maskawa matrix. So, in these expressions $K_{ud}$ is an entry of the Kobayashi--Maskawa matrix diagonal, whose labels are $u$ (standing for a zero--mode up--type quark) and $d$ (representing a zero--mode down--type quark).


Within the heavy--compactification scenario, the Passarino--Velman scalar functions involved in the results can be easily solved through the Feynman--parameters technique to obtain expansions suppressed by powers of the compactification scale. Besides yielding simpler forms of the contributions, this procedure allows one to perform the infinite KK sums in an exact way. The terms comprised by the resulting expressions are all suppressed by powers of the compactification scale. The higher the power of the compactification scale in a given term, the greater the suppression of its contributions, so that we keep terms involving factors $1/M_{\rm c}^2$ and disregard those involving higher powers of the compactification scale. With this in mind, we write the $\Delta\kappa_V^{\rm GNBC}$ and $\Delta Q^{\rm GNBC}_V$ form factors as
\begin{eqnarray}
\label{skg}
\Delta\kappa_V^{\rm GNBC}&=&\frac{1}{M_{\rm c}^2}\frac{\alpha\pi}{432s_{\rm w}^2 \left(Q^2-4 m_{W^{(0)}}^2\right){}^2}\Bigg[-48
   \left(Q^2\right)^3
\nonumber \\ &&
   +m_{W^{(0)}}^2 \left(2 m_{W^{(0)}} \left(-646 Q^2 m_{W^{(0)}}+504m_{W^{(0)}}^3-45 \left(Q^2\right)^{3/2}\right)+527 \left(Q^2\right)^2\right)\bigg]+{\cal O}(1/M_{\rm c}^4),
\\ \nonumber \\
\Delta Q_V^{\rm GNBC}&=&\frac{1}{M_{\rm c}^2}\frac{5 \alpha\pi m_{W^{(0)}}^2}{108s_{\rm w}^2 \sqrt{Q^2}\left(Q^2-4 m_{W^{(0)}}^2\right){}^2}\bigg[-3 \left( Q^2 \right)^{5/2}
\nonumber \\ &&
+m_{W^{(0)}} \left(2
   m_{W^{(0)}}+\sqrt{Q^2}\right) \left(m_{W^{(0)}} \left(-8
   \sqrt{Q^2} m_{W^{(0)}}+6 m_{W^{(0)}}^2+Q^2\right)+3
   \left( Q^2 \right)^{3/2}\right)\bigg]+{\cal O}(1/M_{\rm c}^4),
\end{eqnarray}
where $\alpha$ denotes the fine--structure constant. The $\Delta\kappa_V^{\rm GS}$ and $\Delta Q_V^{\rm GS}$ NGC are given by
\begin{eqnarray}
\Delta\kappa^{\rm GS}_V&=&\frac{1}{M_{\rm c}^2}\frac{\alpha\pi\left(2 s_{\rm w}^2-5\right)\eta_Vm_{W^{(0)}}^2}{288s_{\rm w}^2\left(s_{\rm w}^2-1\right)}+{\cal O}(1/M_{\rm c}^4),
\\ \nonumber \\
\Delta Q_V^{\rm GS}&=&0,
\end{eqnarray}
for which we have defined
\begin{equation}
\eta_V=\eta_{(\gamma^{(0)},Z^{(0)})}=\left( 1,1-\frac{1}{2c_{\rm w}^2} \right).
\end{equation}
The contributions from KK scalar--loops diagrams, $\Delta\kappa_V^{\rm S}$ and $\Delta Q_V^{\rm S}$, can be expressed as
\begin{eqnarray}
\Delta\kappa^{\rm S}_V&=&{\cal O}(1/M_{\rm c}^4),
\\ \nonumber \\
\Delta Q^{\rm S}_V&=&{\cal O}(1/M_{\rm c}^4).
\end{eqnarray}
The latest NGC are the fermionic ones, for which we have separated the cases $V=\gamma^{(0)}$ and $V=Z^{(0)}$. The corresponding expressions are
\begin{eqnarray}
\label{kaf}
\Delta\kappa_{\gamma^{(0)}}^{\rm F}&=&{\cal O}(1/M^4_{\rm c}),
\\ \nonumber \\
\Delta Q_{\gamma^{(0)}}^{\rm F}&=&0,
\\ \nonumber \\
\Delta\kappa^{\rm F}_{Z^{(0)}}&=&\frac{1}{M_{\rm c}^2}\frac{\alpha\pi  Q^2 m_{W^{(0)}}^2 }{72  s_{\rm w}^2
   \left(Q^2-4 m_{W^{(0)}}^2\right){}^2}\left(9 \sqrt{Q^2} m_{W^{(0)}}+2
   m_{W^{(0)}}^2-8 Q^2\right)+{\cal O}(1/M_{\rm c}^4),
\\ \nonumber \\
\label{sQzf}
\Delta Q^{\rm F}_{Z^{(0)}}&=&\frac{1}{M_{\rm c}^2}\frac{\alpha\pi m_{W^{(0)}}^2}{36 \sqrt{Q^2} s_{\rm w}^2 \left(Q^2-4 m_{W^{(0)}}^2\right){}^2} \Big(3 (Q^2)^{5/2}
\nonumber \\ &&
-m_{W^{(0)}} \left(2
   m_{W^{(0)}}+\sqrt{Q^2}\right) \left(m_{W^{(0)}} \left(-8 \sqrt{Q^2} m_{W^{(0)}}+6
   m_{W^{(0)}}^2+Q^2\right)+3 (Q^2)^{3/2}\right)\Big)+{\cal O}(1/M_{\rm c}^4).
\end{eqnarray}

\subsection{Discussion}
The presence of couplings with inverse--mass dimensions in extra--dimensional models indicates that these physical descriptions are not renormalizable. In the context of UED, it was recently shown~\cite{NT1,NT2} that, in spite of this general property, one--loop contributions calculated from models with only one extra dimension are renormalizable. Nonrenormalizability of extra--dimensional theories is not an issue, as it only means that they do not model fundamental descriptions, but are valid below certain energy scale that marks the beginning of a new physical picture. However, the possibility of obtaining finite results, independent of higher--energy scales, is a nice feature that is indeed observed in the contributions to the gauge coupling $WWV$ that we just exhibited. In general, the way through which the nonrenormalizability of extra--dimensional theories manifests, after compactification and integration of the extra dimensions, are the KK sums. To this respect, the final step in the derivation of Eqs. (\ref{skg}) to (\ref{sQzf}) consisted in performing the KK infinite sums, which were found to be known convergent series, namely, Riemann zeta functions. This proves that presumable sources of ultraviolet divergences associated to discrete sums are eliminated. In the results written in terms of Passarino--Veltman scalar functions the disposition of the two--point scalar functions, $B_0$, shows that continuous ultraviolet divergences are exactly cancelled. This feature is even more explicit in the series suppressed by powers of the compactification scale, for not even a track of divergences remain. Thus our results show explicitly that the ultraviolet divergences generated by continuous sums also vanish, so that the final expressions are finite and cutoff independent. Our low--energy physical picture is the SM, which is a renormalizable description where a linear realization of electroweak symmetry breaking takes place. This context sets the conditions under which the decoupling theorem~\cite{AC} is fulfilled. Note that all our results consistently decouple in the limit $M_{\rm c}\to\infty$, which can be better appreciated in Eqs. (\ref{skg}) to (\ref{sQzf}), since the suppression provided by the compactification scale makes it explicit in them. We divided the contributions to the TGC's in order to emphasize their behavior with respect to gauge symmetry dictated by the $SU_4(2)_{\rm L}$ group. The general parametrization of the new--physics effects on the $WWV$ interactions arises, from the viewpoint of effective field theory, from gauge invariant effective terms, both renormalizable and nonrenormalizable, whose building blocks are the SM gauge symmetry and dynamic variables. The physical states associated to the photon and $Z$ boson are then innately related to each other through the $W^{(0)3}_\mu$ field, due to electroweak gauge symmetry. It is through such link that the new--physics contributions to the TGC's $WWV$ parametrization that are governed by the $SU_4(2)_{\rm L}$ gauge group are expected to distinguish the cases $V=\gamma^{(0)}$ and $V=Z^{(0)}$ only through the couplings $g_V$. In the case of the GNBC, notice that there is no difference between the form factors corresponding to the case $V=\gamma^{(0)}$ and those arising for $V=Z^{(0)}$, which indicates, according to the above discussion, that $SU_4(2)_{\rm L}$ gauge symmetry is maintained at the one--loop level by such contributions. Contrastingly, the NGC distinguish the cases $V=\gamma^{(0)}$ and $V=Z^{(0)}$, which leads to the conclusion that they are not governed by the $SU_4(2)_{\rm L}$ gauge group. The reason behind such different comportment lies on the role played by electroweak gauge symmetry, which is innocuous to the sources of GNBC, but strikes the couplings of KK excited and zero modes originated in the five--dimensional scalar and fermionic sectors, whose subsequent yields are the NGC.
\begin{figure}[ht]
\center
\includegraphics[width=8.3cm]{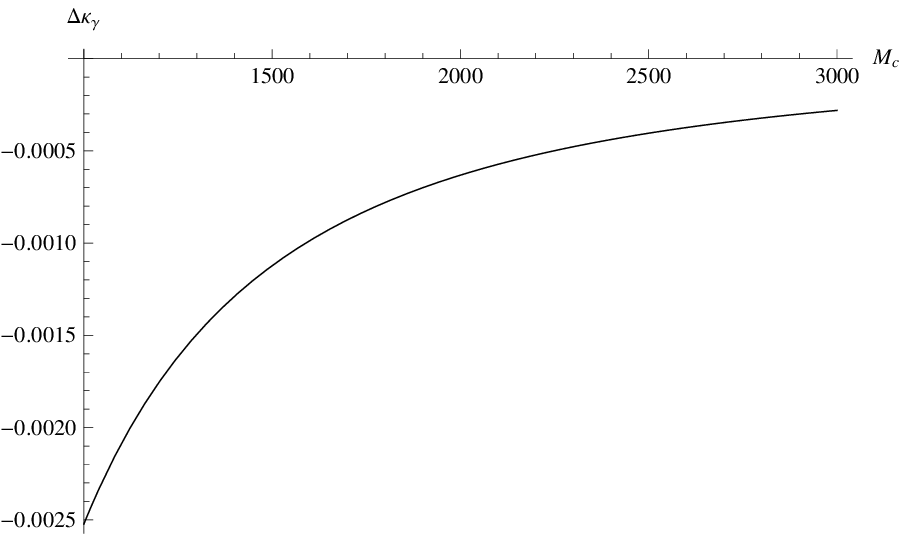}\hspace{1cm}
\includegraphics[width=8.3cm]{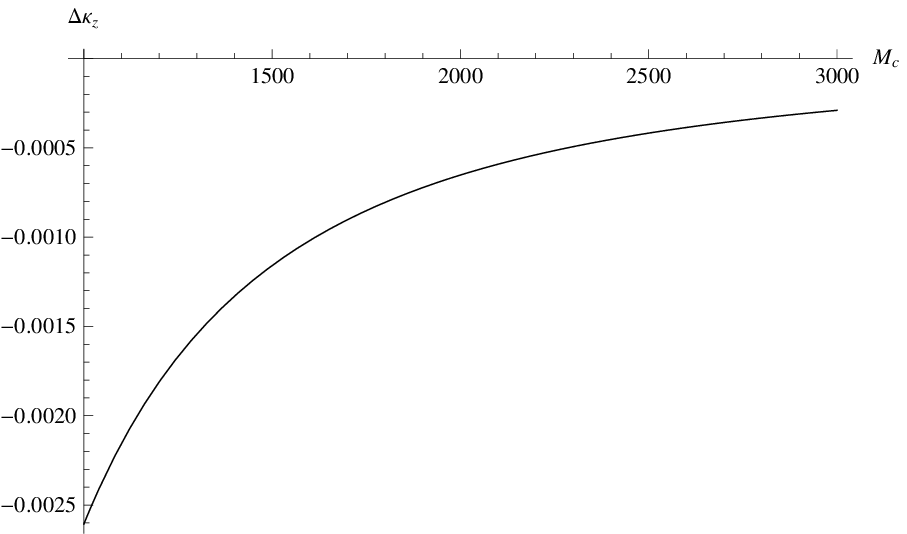}
\\ \vspace{1cm}
\includegraphics[width=8.3cm]{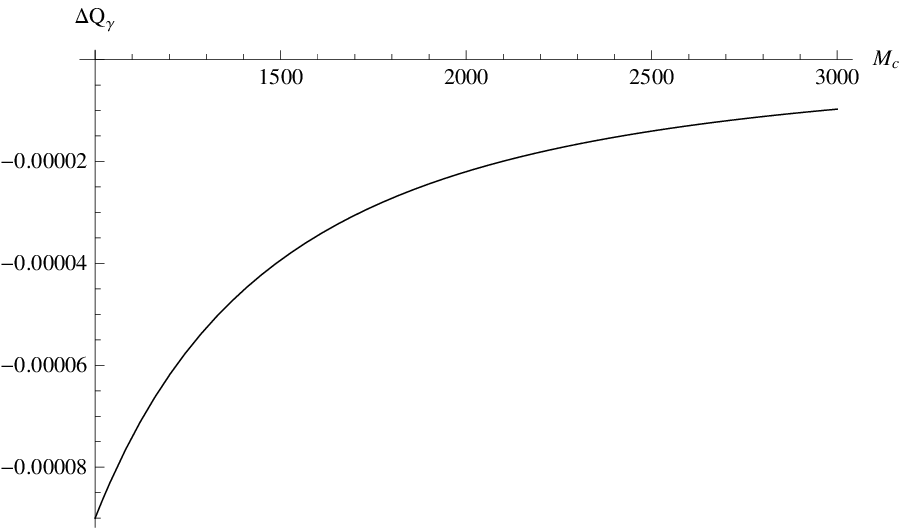}\hspace{1cm}
\includegraphics[width=8.3cm]{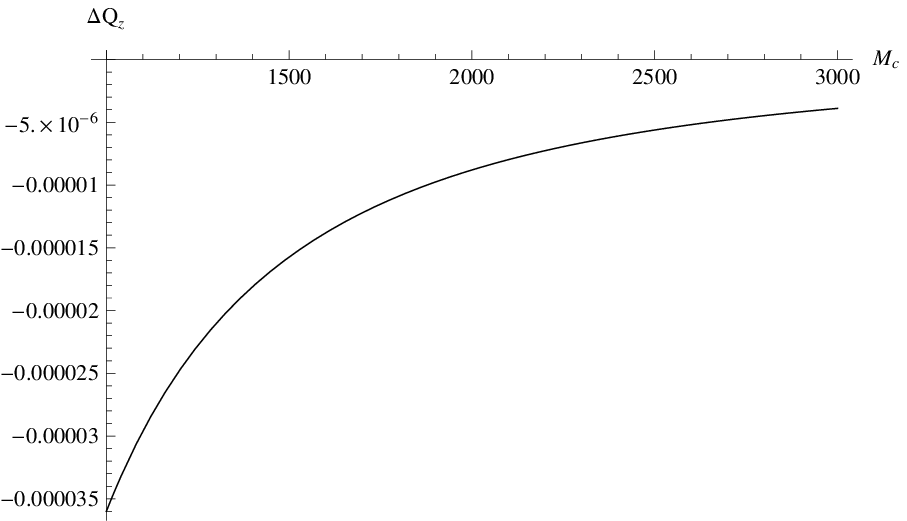}
\caption{\label{plots} Behavior of the $\Delta\kappa_V$ and $\Delta Q_V$ with respect to compactification scale, $M_{\rm c}$, at $\sqrt{Q^2}=500$~GeV. All extra--dimensional contributions decouple in the limit $M_{\rm c}\to\infty$.}
\end{figure}

The determination of the existence of UED could be achieved at the LHC through direct production of the lightest KK particle, which is neutral and, as a consequence of KK--parity conservation, stable, so that its production at colliders would result in generic missing--energy signals. The detection of such KK excited mode would be challenging, for the high degeneracy of the mass spectrum of KK excited modes of the same KK level leads to small missing--energy signals. Direct production of second--level KK states, which would be a way to reliably determine the UED nature of the observed new physics effects, could be out of the reach of the LHC, as KK excited modes are exclusively pair--produced at tree level. The study of virtual effects, such as those associated to TGC's, could then provide useful tools for the search and determination of UED. The TGC's $WWV$ can be studied through the LHC and the ILC as well, but it has been pointed out~\cite{ilcandlhc} that the sensitivity of the latter to this gauge interaction is much better than that corresponding to the former. While the LHC is expected~\cite{ilcandlhc} to constrain $\Delta Q_V$ up to ${\cal O}(10^{-3})$ and to set a bound as restrictive as ${\cal O}(10^{-2})$ on $\Delta\kappa_V$, the sensitivity of ILC should reach~\cite{ilcandlhc} bounds of ${\cal O}(10^{-4})$ on all these parameters. The SM one--loop contributions to the $\Delta\kappa_{\gamma^{(0)}}$ and $\Delta Q_{\gamma^{(0)}}$ form factors have been already calculated by employing conventional~\cite{AKLPS} and unconventional~\cite{PP} gauge--fixing approaches. The gauge--invariant scheme followed in Ref.~\cite{PP} leaded to a SM contribution to $\Delta\kappa_{\gamma^{(0)}}$ varying from $10^{-3}$ to $10^{-4}$ for energies in the range $200$~GeV$<\sqrt{Q^2}<1000$~GeV, while, for the same energy values, the $\Delta Q_{\gamma^{(0)}}$ form factor is found~\cite{AKLPS} to lie between $10^{-4}$ and $10^{-5}$. In Fig. (\ref{plots}), one can appreciate the behavior of the extra--dimensional contributions to all form factors as larger compactification energy scales are considered within a scenario in which $\sqrt{Q^2}=500$~GeV. Also, in Table~\ref{numtab},
\begin{table}[ht]
\centering
\begin{tabular}{|c|c|c|c|c|}
\hline
&$\Delta\kappa_{\gamma^{(0)}}$ & $\Delta\kappa_{Z^{(0)}}$ & $\Delta Q_{\gamma^{(0)}}$ & $\Delta Q_{Z^{(0)}}$
\\
\hline
$M_{\rm c}=$1000~GeV&$-2.5\times10^{-3}$&$-2.6\times10^{-3}$&$-9.0\times10^{-5}$&$-3.6\times10^{-5}$
\\
\hline
$M_{\rm c}=$2000~GeV&$-6.3\times10^{-4}$&$-6.5\times10^{-4}$&$-2.2\times10^{-5}$&$-8.8\times10^{-6}$

\\
\hline
$M_{\rm c}=$3000~GeV&$-2.8\times10^{-4}$&$-2.9\times10^{-4}$&$-9.7\times10^{-6}$&$-3.9\times10^{-6}$
\\
\hline
\end{tabular}
\caption{\label{numtab} Values of the $\Delta\kappa_V$ and $\Delta Q_V$ parameters for fixed $\sqrt{Q^2}=500$~GeV.}
\end{table}
we provide a summary of numerical results on all these parameters for $\sqrt{Q^2}=500$~GeV combined with compactification scales $M_{\rm c}=1000$~GeV, $M_{\rm c}=2000$~GeV, and $M_{\rm c}=3000$~GeV. These numerical estimations were all made by employing the expansions of the form factors with terms suppressed by powers of the compactification scale $M_{\rm c}$. Our results indicate that the $\Delta\kappa_V$ range from $\Delta\kappa_V\sim-10^{-3}$ to $-10^{-4}$, for compactification scales within $M_{\rm c}=1$~TeV and $3$~TeV, which means that ILC would be sensitive at $E_{\rm CM}=500$~GeV in these scenarios. In the case of $\Delta\kappa_{\gamma^{(0)}}$, note that the one--loop extra--dimensional contributions are about the same order of magnitude than those from the SM. On the other hand, we find the $\Delta Q_V$ parameters to be out of the reach of ILC, as for $\sqrt{Q^2}=500$~GeV we estimate their values to be between ${\cal O}(10^{-5})$ and ${\cal O}(10^{-6})$ for $1\hspace{0.05cm}{\rm TeV}< M_{\rm c}<3\hspace{0.05cm}{\rm TeV}$. This loop contribution is minor than the one emerging from the SM by about one order of magnitude.

As we performed all numerical estimations in a heavy--compactification scenario, for which the mass spectrum of the KK excited modes was taken to be degenerate, one might wonder whether the impact of the top quark mass on the TGC's could have been minimized by this assumption and actually play an important role in the non--degenerate framework. We evaluated the contributions involving the KK excited modes of the top quark in an exact way, that is, by taking their mass as given by $m^2_{t^{(m)}}=m^2_{t^{(0)}}+m_n^2$, and found that the total fermionic contributions to the $\Delta Q_V$ form factors do not experience appreciable changes, while the corresponding contributions to the $\Delta\kappa_V$ are highly sensitive to quarks' heavy masses. We have verified that the same pattern occurs in the one--loop fermionic contributions from the SM, which exceed the importance of the extra--dimensional contributions by about one or two orders of magnitude in all cases. In spite of the enhancement observed in this calculation, the total numerical results are not notoriously modified, so that the leading extra--dimensional contributions to these from factors are still engendered by the gauge sector of the five--dimensional high--energy description. The only case in which the order of magnitude of such contributions is matched by another source is the $\Delta Q_{Z^{(0)}}$ form factor, which also receives a large contribution, between $\sim10^{-5}$ and $\sim10^{-6}$, from the extra--dimensional fermionic sector, although, as already mentioned, such form factor is not much sensitive to the top quark mass and the extra--dimensional pure--gauge contribution is still the most important. In all cases, contributions from the five--dimensional gauge sector are negative, so that the total contributions to all form factors are negative as well \footnote{The SM one--loop contributions to the $WW\gamma$ vertex reported in Ref.~\cite{BGL} consistently have the same signs as our results, as expected since the SM couplings and those involving KK excited modes are similar. At the first glance, there seems to be discrepancies between the signs of our results and the ones derived in this reference, but all differences are caused by the conventions taken in that work and the ones used in the present paper for the general $WWV$ parametrization.}. This can be observed in the plots of Fig.~(\ref{plots}) and in the data of Table~\ref{numtab}. There exists the possibility of including effects on the TGC's $WWV$ from an extra--dimensional effective theory parametrizing the impact of the high--energy description lying beyond the cutoff of the 5DSM. Under reasonable assumptions, it has been estimated~\cite{NT2} that the effects of such presumable higher--energy theory amount to a small percentage of those generated by the SM in five dimensions, for which we have omitted a comparison in the present paper.

\section{Conclusions}
\label{conc}
The main purpose of the present paper was the study of the one--loop effects conceived by an extra--dimensional generalization of the SM on the parameters characterizing the charged TGC's $WWV$. Motivated by the importance of one--loop calculations in extra--dimensional extensions of the SM in which the extra dimensions are assumed to be universal, we considered the case of one UED. The interesting gauge structure of the KK theory obtained after compactification of the extra dimension was emphasized. Two sorts of gauge transformations arise at the four dimensional level, one of which contains the low--energy SM gauge symmetry, whereas the other corresponds to the gauge nature of some heavy degrees of freedom, that is, KK excited modes. In order to perform phenomenological calculations, it was adequate to remove the gauge symmetry characterized by the latter set of gauge transformations, which we accomplished through an interesting covariant gauge--fixing procedure that gave rise to a quantum theory that is still invariant under the electroweak gauge group. By virtue of the importance of $W^+W^-$ production in linear colliders, which is a window to study TGC's, we took the external SM $W$ bosons on--shell, but kept the SM neutral gauge boson off--shell. We then performed the calculation, in the Feynman--'t Hooft gauge, in a context in which the compactification scale was supposed to be large. Within this heavy--compactification scenario, we set the masses of the KK excited modes to be all the same, for such a degenerate mass spectrum yields valuable simplifications. All ultraviolet divergences associated to continuous sums were found to vanish and the results behaved in a decoupling way. The heavy--compactification scenario then made it possible to express all results as series with terms suppressed by the powers of the compactification scale, where it was emphasized that divergences associated to discrete infinite KK sums as well lead to finite results, from which we concluded that the final expressions are finite and independent of higher--energy scales. Each contribution to the $WWV$ form factors was subdivided into two parts, one of which is unaffected by electroweak symmetry breaking and consequently shows manifest invariance with respect to the $SU_4(2)_{\rm L}$ gauge group. The other part in each contribution, emerged from the extra--dimensional scalar and fermionic sectors, did not exhibit such a comportment, which is related to the sensitivity of its extra--dimensional originating source to spontaneous breaking of the electroweak group. Our numerical estimations produced dominating contributions from the extra--dimensional gauge sector, which were not even exceeded by those associated to fermionic excited modes. Form factors were found to be comparable to the SM one--loop corrections and well within the reach of a linear collider with a center--of--mass energy amounting to 500~GeV and compactification scales ranging from $1000$~GeV to $3000$~GeV.

\acknowledgments{We acknowledge financial support from CONACYT (M\'exico). J. M., J. J. T. and E. S. T. acknowledge financial support from SNI (M\'exico). H. --N. S and J. J. T. acknowledge financial support from VIEP--BUAP.}

\appendix

\section{Lagrangians contributing to $WWV$ at the one--loop level}
\label{AppL}
In this section, we provide the Lagrangian terms contributing to the $WWV$ vertex at the one-loop level.
\subsection*{Pure--gauge contributions}
The part of the KK Lagrangian generated by the extra--dimensional gauge sector that yields one--loop contributions to the TGC's $WWV$ can be expressed as
\begin{eqnarray}
{\cal L}^{\rm 1-loop}_{\rm G}&=&{\cal L}_{W^{(0)3}W^{(n)-}W^{(n)+}}+{\cal L}_{W^{(0)\mp}W^{(n)\pm}W^{(n)3}}+{\cal L}_{W^{(0)-}W^{(0)+}W^{(n)3}W^{(n)3}}+{\cal L}_{W^{(0)\mp}W^{(n)\pm}W^{(0)3}W^{(n)3}}
\nonumber \\ \nonumber &&
+{\cal L}_{W^{(0)3}W^{(n)-}_5W^{(n)+}_5}+{\cal L}_{W^{(0)\mp}W^{(n)\pm}_5W^{(n)3}_5}+{\cal L}_{W^{(0)-}W^{(0)+}W^{(n)3}_5W^{(n)3}_5}+{\cal L}_{W^{(0)\mp}W^{(n)\pm}_5W^{(0)3}W^{(n)3}_5}
\\ &&
{\cal L}_{W^{(0)}C^{(n)}\bar{C}^{(n)}}+{\cal L}_{W^{(0)}W^{(0)}C^{(n)}\bar{C}^{(n)}},
\end{eqnarray}
where
\begin{eqnarray}
{\cal L}_{W^{(0)3}W^{(n)-}W^{(n)+}}&=&-ig\bigg[\left(W^{(n)+}_{\mu \nu}W^{(n)-\nu} -W^{(n)-}_{\mu \nu}W^{(n)+\nu} \right) W^{(0)3\mu}+W^{(0)3}_{\mu \nu}W^{(n)-\mu}W^{(n)+\nu}
\nonumber \\ &&
-\frac{1}{\xi}W^{(0)3}_\mu\left( W^{(n)+\mu}\partial_\nu W^{(n)-\nu}-W^{(n)-\mu}\partial_\nu W^{(n)+\nu} \right)
\bigg],
\end{eqnarray}
\begin{eqnarray}
{\cal L}_{W^{(0)\mp}W^{(n)\pm}W^{(n)3}}&=&-ig\bigg[\left(W^{(n)-}_{\mu \nu}W^{(n)3\nu}-W^{(n)3}_{\mu \nu}W^{(n)-\nu}  \right)W^{(0)+\mu}+W^{(0)+}_{\mu \nu}W^{(n)3\mu}W^{(n)-\nu}
\nonumber \\ &&
-\left(W^{(n)+}_{\mu \nu}W^{(n)3\nu}-W^{(n)3}_{\mu \nu}W^{(n)+\nu}  \right)W^{(0)-\mu}
-W^{(0)-}_{\mu \nu}W^{(n)3\mu}W^{(n)+\nu}
\nonumber
\\ \nonumber &&
-\frac{1}{\xi}W^{(0)+}_\mu\left( W^{(n)-\mu}\partial_\nu W^{(n)3\nu}-W^{(n)3\mu}\partial_\nu W^{(n)-\nu} \right)
\\ &&
+\frac{1}{\xi}W^{(0)-}_\mu\left( W^{(n)+\mu}\partial_\nu W^{(n)3\nu}-W^{(n)3\mu}\partial_\nu W^{(n)+\nu} \right)\bigg],
\end{eqnarray}
\begin{eqnarray}
{\cal L}_{W^{(0)-}W^{(0)+}W^{(n)-}W^{(n)+}}&=&\frac{g^2}{2}
\bigg[ \left( W^{(0)+}_\mu W^{(n)-}_\nu-W^{(0)+}_\nu W^{(n)-}_\mu \right)\left( W^{(0)-\nu}W^{(n)+\mu}-W^{(0)-\mu}W^{(n)\nu} \right)
\nonumber \\ \nonumber &&
+\left( W^{(n)-}_\mu W^{(n)+}_\nu-W^{(n)-}_\nu W^{(n)+}_\mu \right)\left( W^{(0)-\mu}W^{(0)+\nu}-W^{(0)-\nu}W^{(0)+\mu} \right)
\\ &&
-\frac{2}{\xi}W^{(0)+}_\mu W^{(0)-}_\nu W^{(n)-\mu}W^{(n)+\nu}
\bigg],
\end{eqnarray}
\begin{eqnarray}
{\cal L}_{W^{(0)\mp}W^{(n)\pm}W^{(0)3}W^{(n)3}}&=&-\frac{g^2}{2}\bigg[\left(W^{(n)-}_\mu W^{(0)3}_\nu -W^{(n)-}_\nu W^{(0)3}_\mu\right)\left(W^{(0)+\mu}W^{(n)3\nu}-W^{(0)+\nu}W^{(n)3\mu}\right)
\nonumber \\ \nonumber &&
+\left(W^{(0)+}_\mu W^{(0)3}_\nu-W^{(0)+}_\nu W^{(0)3}_\mu \right)\left(W^{(n)-\mu}W^{(n)3\nu}-W^{(n)-\nu}W^{(n)3\mu} \right)
\\  &&
-\frac{2}{\xi}W^{(n)3\mu}W^{(0)3}_\nu W^{(n)-\nu}W^{(0)+}_\mu
+{\rm H.\, c.}\bigg],
\end{eqnarray}
\begin{eqnarray}
{\cal L}_{W^{(0)3}W^{(n)-}_5W^{(n)+}_5}&=&igW^{(0)3}_\mu\left( W^{(n)-}_5\partial^\mu W^{(n)+}_5-W^{(n)+}_5\partial^\mu W^{(n)-}_5 \right),
\end{eqnarray}
\begin{eqnarray}
{\cal L}_{W^{(0)\mp}W^{(n)\pm}_5W^{(n)3}_5}&=&-ig\Big[ W^{(0)-}_\mu\left( W^{(n)3}_5\partial^\mu W^{(n)+}_5-W^{(n)+}_5\partial^\mu W^{(n)3}_5 \right)
\nonumber \\  &&
-W^{(0)+}_\mu\left( W^{(n)3}_5\partial^\mu W^{(n)-}_5-W^{(n)-}_5\partial^\mu W^{(n)3}_5 \right) \Big],
\\ \nonumber \\
{\cal L}_{W^{(0)-}W^{(0)+}W^{(n)-}_5W^{(n)+}_5}&=&g^2W^{(0)-}_\mu W^{(0)+\mu}W^{(n)-}_5W^{(n)+}_5,
\\ \nonumber \\
{\cal L}_{W^{(0)\mp}W^{(n)\pm}_5W^{(0)3}W^{(n)3}_5}&=&-g^2W^{(0)3}_\mu W^{(n)3}_5\left( W^{(0)+\mu}W^{(n)-}_5+W^{(0)-\mu} W^{(n)+}_5 \right),
\end{eqnarray}
\begin{eqnarray}
{\cal L}_{W^{(0)}C^{(n)}\bar{C}^{(n)}}&=&-ig\Big[ W^{(0)3}_\mu\left( C^{(n)+}\partial^\mu\bar{C}^{(n)-}-\partial^\mu C^{(n)+}\bar{C}^{(n)-}-C^{(n)-}\partial^\mu\bar{C}^{(n)+}+\partial^\mu C^{(n-)}\bar{C}^{(n)+} \right)
\nonumber \\ \nonumber &&
-W^{(0)+}_\mu\left( C^{(n)3}\partial^\mu\bar{C}^{(n)-}-\partial^\mu C^{(n)3}\bar{C}^{(n)-}+\partial^\mu C^{(n)-}\bar{C}^{(n)3}-C^{(n)-}\partial^\mu\bar{C}^{(n)3} \right)
\\  &&
+W^{(0)-}_\mu\left( C^{(n)3}\partial^\mu\bar{C}^{(n)+}-\partial^\mu C^{(n)3}\bar{C}^{(n)+}+\partial^\mu C^{(n)+}\bar{C}^{(n)3}-C^{(n)+}\partial^\mu\bar{C}^{(n)3} \right)
\Big],
\\ \nonumber \\
{\cal L}_{W^{(0)}W^{(0)}C^{(n)}\bar{C}^{(n)}}&=&g^2\Big[ W^{(0)-}_\mu W^{(0)+\mu}\left( C^{(n)+}\bar{C}^{(n)-}+C^{(n)-}\bar{C}^{(n)+}+2C^{(n)3}\bar{C}^{(n)3} \right)
\nonumber \\ \nonumber &&
-W^{(0)+}_\mu W^{(0)+\mu}C^{(n)-}\bar{C}^{(n)-}-W^{(0)-}_\mu W^{(0)-\mu}C^{(n)+}\bar{C}^{(n)+}-W^{(0)3}_\mu W^{(0)3\mu}C^{(n)3}\bar{C}^{(n)3}
\\ &&
-W^{(0)3}_\mu\Big( W^{(0)+\mu}\left( C^{(n)-}\bar{C}^{(n)3}+C^{(n)3}\bar{C}^{(n)-} \right)
\nonumber \\ &&
+W^{(0)-\mu}\left( C^{(n)+}\bar{C}^{(n)3}+C^{(n)3}\bar{C}^{(n)+} \right) \Big)
\Big],
\end{eqnarray}
for which we have defined
\begin{eqnarray}
C^{(n)\pm}&=&\frac{1}{\sqrt{2}}\left( C^{(n)1}\mp iC^{(n)2} \right),
\\ \nonumber \\
\bar{C}^{(n)\pm}&=&\frac{1}{\sqrt{2}}\left( \bar{C}^{(n)1}\mp i\bar{C}^{(n)2} \right).
\end{eqnarray}

\subsection*{KK--excited--scalar contributions}
The contributing term containing couplings of scalar KK excitations to gauge KK zero and excited modes read
\begin{equation}
{\cal L}^{\rm 1-loop}_{\rm S}={\cal L}_{W^{(0)}({\rm KK})({\rm KK})}+{\cal L}_{Z^{(0)}({\rm KK})({\rm KK})}+{\cal L}_{A^{(0)}({\rm KK})({\rm KK})}+{\cal L}_{W^{(0)}W^{(0)}({\rm KK})({\rm KK})},
\end{equation}
with
\begin{eqnarray}
{\cal L}_{W^{(0)}({\rm KK})({\rm KK})}&=&gm_{W^{(0)}}H^{(n)}\left(W^{(0)+}_\mu W^{(n)-\mu}+W^{(0)-}_\mu W^{(n)+\mu}\right)  \nonumber \\ &&
+\frac{g}{2}\left(2c_Ws_\alpha s_\beta+c_\alpha c_\beta \right)
\Big[h^{(n)}\left(W^{(0)+}_\mu \partial^\mu H^{(n)-}+W^{(0)-}_\mu \partial^\mu H^{(n)+}\right)\nonumber \\
&&-\left(W^{(0)+}_\mu H^{(n)-}+W^{(0)-}_\mu H^{(n)+}\right)\partial^\mu h^{(n)}\Big]
+\frac{3}{2}igc_Wc_\beta m_{Z^{(0)}}h^{(n)}\left(W^{(0)-\mu}W^{(n)+}_\mu-W^{(0)+\mu}W^{(n)-}_\mu\right)
\nonumber \\
&&-\frac{igc_\alpha}{2}\Big[H^{(n)}\left(W^{(0)+}_\mu \partial^\mu H^{(n)-}-W^{(0)-}_\mu \partial^\mu H^{(n)+}\right)-\left(W^{(0)+}_\mu H^{(n)-}-W^{(0)-}_\mu H^{(n)+}\right)\partial^\mu H^{(n)}\Big]
\nonumber \\ &&
+gc_\alpha \left[m_{W^{(0)}}s_WA^{(n)}_\mu-m_{Z^{(0)}}\left(\frac{1+2s^2_W}{4}\right)Z^{(n)}_\mu \right]\left(W^{(0)+}_\mu H^{(n)-}+W^{(0)-}_\mu H^{(n)+}\right),
\end{eqnarray}
\begin{eqnarray}
{\cal L}_{Z^{(0)}({\rm KK})({\rm KK})}&=&-igc_W\left(1-\frac{c^2_\alpha}{2c^2_W}\right) Z^{(0)}_\mu \Big(H^{(n)+}\partial^\mu H^{(n)-}-H^{(n)-}\partial^\mu H^{(n)+}\Big)
\nonumber \\ &&
-gc_\alpha m_{Z^{(0)}}Z^{(0)}_\mu \left(H^{(n)-}W^{(n)+\mu}+H^{(n)+}W^{(n)-\mu}\right)
\nonumber \\  &&
+\frac{gc_\beta}{2c_W}Z^{(0)}_\mu \left(h^{(n)}\partial^\mu H^{(n)}-H^{(n)}\partial^\mu h^{(n)}\right)+\frac{gm_{Z^{(0)}}}{c_W}Z^{(0)}_\mu Z^{(n)\mu}H^{(n)},
\\ \nonumber \\
{\cal L}_{A^{(0)}({\rm KK})({\rm KK})}&=&-ieA^{(0)}_\mu \left(H^{(n)+}\partial^\mu H^{(n)-}-H^{(n)-}\partial^\mu H^{(n)+}\right),
\end{eqnarray}
\begin{eqnarray}
{\cal L}_{W^{(0)}W^{(0)}({\rm KK})({\rm KK})}&=&\frac{g^2}{4}\Big\{W^{(0)-}_\mu W^{(0)+\mu}\Big[2\left(1+s^2_\alpha \right)H^{(n)-} H^{(n)+}
+H^{(n)}H^{(n)} +\left(c^2_\beta+4c^2_Ws^2_\beta \right)h^{(n)}h^{(n)} \Big]
 \nonumber \\ &&
+2s^2_\alpha \Big(W^{(0)+}_\mu W^{(0)+\mu}H^{(n)-}H^{(n)-}+W^{(0)-}_\mu W^{(0)-\mu}H^{(n)+}H^{(n)+} \Big) \Big\}\, ,
\end{eqnarray}

\subsection*{Fermionic contributions}
The contributions generated by fermionic KK excited modes coupling to a zero--mode gauge boson come from the Lagrangian
\begin{equation}
{\cal L}^{\rm 1-loop}_{\rm F}={\cal L}_{W^{(0)}\nu^{(n)}e^{(n)}}+{\cal L}_{Z^{(0)}e^{(n)}e^{(n)}}+{\cal L}_{A^{(0)}e^{(n)}e^{(n)}}+{\cal L}_{W^{(0)}u^{(n)}d^{(n)}}+{\cal L}_{Z^{(0)}q^{(n)}q^{(n)}}+{\cal L}_{A^{(0)}q^{(n)}q^{(n)}},
\end{equation}
where we have defined the leptons--gauge interaction terms
\begin{eqnarray}
{\cal L}_{W^{(0)}\nu^{(n)}e^{(n)}}&=&\frac{g}{\sqrt{2}} \left[\cos\frac{\alpha^{(n)}_e}{2}\left(\bar{N}^{(n)}\gamma^\mu E^{(n)}\right)
-\sin\frac{\alpha^{(n)}_e}{2}\left(\bar{N}^{(n)}\gamma^\mu \gamma^5 \hat{E}^{(n)}\right)\right]W^{(0)+}_\mu
+{\rm H.\, c.}\, ,
\end{eqnarray}
\begin{eqnarray}
{\cal L}_{Z^{(0)}e^{(n)}e^{(n)}}&=&\frac{g}{2c_W} \Bigg[\bar{N}^{(n)}\gamma^\mu N^{(n)}+\left(\bar{E}^{(n)} \, \, \, \bar{\hat{E}}^{(n)}\right)\gamma^\mu \left(\begin{array}{ccc}
Z_{EE} & Z_{E\hat{E}} \\
\, \\
Z_{\hat{E}E} & Z_{\hat{E}\hat{E}}
\end{array}\right) \left(\begin{array}{ccc}
E^{(n)} \\
\, \\
\hat{E}^{(n)}
\end{array}\right)\Bigg]Z^{(0)}_\mu \, ,
\end{eqnarray}
\begin{equation}
{\cal L}_{A^{(0)}e^{(n)}e^{(n)}}=-e  \left(\bar{E}^{(n)}\gamma^\mu E^{(n)}+\bar{\hat{E}}^{(n)}\gamma^\mu \hat{E}^{(n)} \right)A^{(0)}_\mu \, .
\end{equation}
along with the definitions
\begin{eqnarray}
Z_{EE}&=&\cos^2\frac{n_{e^{(n)}}}{2}-2s^2_W \, , \\
Z_{\hat{E}\hat{E}}&=&\sin^2\frac{n_{e^{(n)}}}{2}-2s^2_W \, , \\
Z_{E\hat{E}}&=&Z_{\hat{E}E}=\sin\frac{n_{e^{(n)}}}{2}\cos\frac{n_{e^{(n)}}}{2}\, .
\end{eqnarray}
On the other hand, the quarks--gauge couplings are produced by
\begin{eqnarray}
{\cal L}_{W^{(0)}u^{(n)}d^{(n)}}&=&\frac{g}{\sqrt{2}}\Bigg[
\left(\bar{U}^{(n)} \, \, \, \bar{\hat{U}}^{(n)}\right)K\gamma^\mu \left(\begin{array}{ccc}
W_{UD} & W_{U\hat{D}} \\
\, \\
W_{\hat{U}D} & W_{\hat{U}\hat{D}}
\end{array}\right) \left(\begin{array}{ccc}
D^{(n)} \\
\, \\
\hat{D}^{(n)}
\end{array}\right)\Bigg]W^{(0)+}_\mu+{\rm H.\, c.} \, ,
\end{eqnarray}
\begin{eqnarray}
{\cal L}_{Z^{(0)}q^{(n)}q^{(n)}}&=&\frac{g}{c_W}Z^{(0)}_\mu\Bigg\{ \left(\bar{U}^{(n)} \, \, \, \bar{\hat{U}}\hspace{0.001cm}^{(n)}\right)\gamma^\mu \left(\begin{array}{ccc}
Z^q_{UU} & Z^q_{U\hat{U}}\gamma^5 \\
\, \\
Z^q_{\hat{U}U}\gamma^5 & Z^q_{\hat{U}\hat{U}}
\end{array}\right) \left(\begin{array}{ccc}
U^{(n)} \\
\, \\
\hat{U}^{(n)}
\end{array}\right)
 \nonumber \\ &&
 +\left(\bar{D}^{(n)} \, \, \, \bar{\hat{D}}\hspace{0.001cm}^{(n)}\right)\gamma^\mu \left(\begin{array}{ccc}
Z^q_{DD} & Z^q_{D\hat{D}}\gamma^5 \\
\, \\
Z^q_{\hat{D}D}\gamma^5 & Z^q_{\hat{D}\hat{D}}
\end{array}\right) \left(\begin{array}{ccc}
D^{(n)} \\
\, \\
\hat{D}^{(n)}
\end{array}\right)
\Bigg\},
\end{eqnarray}
\begin{equation}
{\cal L}_{A^{(0)}q^{(n)}q^{(n)}}=e\sum_{q=u,d,\ldots }Q_q\left(
\bar{q}^{(n)}\gamma^\mu q^{(n)}+\bar{\hat{q}}^{(n)}\gamma^\mu \hat{q}^{(n)}\right)A^{(0)}_\mu \,,
\end{equation}
where
\begin{eqnarray}
W_{UD}&=&\cos \frac{n_{u^{(n)}}}{2}\cos \frac{n_{d^{(n)}}}{2} \, , \\ \nonumber \\
W_{\hat{U}\hat{D}}&=&\sin \frac{n_{u^{(n)}}}{2}\sin \frac{n_{d^{(n)}}}{2} \, , \\ \nonumber \\
W_{U\hat{D}}&=&-\sin \frac{n_{d^{(n)}}}{2}\cos \frac{n_{u^{(n)}}}{2}\, ,\\\nonumber \\
W_{\hat{U}D}&=&-\sin \frac{n_{u^{(n)}}}{2}\cos \frac{n_{d^{(n)}}}{2}\, ,
\\\nonumber \\
Z^q_{UU}&=&\left( \frac{1}{2}-Q_u\right)\cos^2\frac{n_{u^{(n)}}}{2}-Q_u s^2_W\sin^2\frac{n_{u^{(n)}}}{2} \, , \\\nonumber \\
Z^q_{\hat{U}\hat{U}}&=&\left( \frac{1}{2}-Q_u\right)\sin^2\frac{n_{u^{(n)}}}{2}-Q_u s^2_W\cos^2\frac{n_{u^{(n)}}}{2}\, , \\\nonumber \\
Z^q_{U\hat{U}}&=&Z^q_{\hat{U}U}=-\left(\frac{1}{2}-Q_uc^2_W\right)\sin\frac{n_{u^{(n)}}}{2}\cos\frac{n_{u^{(n)}}}{2}\, ,
\\ \nonumber \\
Z^q_{DD}&=&\left( \frac{1}{2}+Q_d\right)\cos^2\frac{n_{d^{(n)}}}{2}+Q_d s^2_W\sin^2\frac{n_{d^{(n)}}}{2} \, , \\\nonumber \\
Z^q_{\hat{D}\hat{D}}&=&\left( \frac{1}{2}+Q_d\right)\sin^2\frac{n_{d^{(n)}}}{2}+Q_d s^2_W\cos^2\frac{n_{d^{(n)}}}{2} \, , \\\nonumber \\
Z^q_{D\hat{D}}&=&Z^q_{\hat{D}D}=-\left(\frac{1}{2}+Q_dc^2_W\right)\sin\frac{n_{d^{(n)}}}{2}\cos\frac{n_{d^{(n)}}}{2} \, .
\end{eqnarray}

\end{document}